\documentclass[aps,pra,twocolumn,superscriptaddress,showpacs]{revtex4} 
\usepackage{graphicx}
\usepackage{amsmath,amssymb,color} 
\def\degree{\mbox{$^\circ$}}
\def\U#1{{%
\def\O{\mbox{O}}
\def\u{\mbox{u}}
\mathcode`\u=\mu
\mathcode`\O=\Omega
\mathrm{#1}}}
\def\ii{{\mathrm{i}}}
\def\ee{{\mathrm{e}}}
\def\dd{{\mathrm{d}}}

\def\ket#1{|\mbox{$#1$}\rangle}

\def\bracket#1{\langle\mbox{$#1$}\rangle}
\def\bracketi#1#2{\langle\mbox{$#1$}|\mbox{$#2$}\rangle}

\def\fracpd#1#2{\frac{\partial#1}{\partial#2}}
\def\sub#1{_{\scriptsize\mbox{#1}}}

\begin{document}
\title{Nonlinear behavior of geometric phases induced by photon pairs}
\author{H. Kobayashi}
\affiliation{Department of Electronic Science and Engineering, Kyoto
University, Kyoto 615-8510, Japan}
\author{Y. Ikeda}
\affiliation{Department of Electronic Science and Engineering, Kyoto
University, Kyoto 615-8510, Japan}
\author{S. Tamate}
\affiliation{Department of Electronic Science and Engineering, Kyoto
University, Kyoto 615-8510, Japan}
\author{T. Nakanishi}
\affiliation{Department of Electronic Science and Engineering, Kyoto
University, Kyoto 615-8510, Japan}
\author{M. Kitano}
\affiliation{Department of Electronic Science and Engineering, Kyoto
University, Kyoto 615-8510, Japan}
\date{\today}

\begin{abstract}
In this study, we observe the nonlinear behavior of the two-photon geometric phase
 for polarization states using time-correlated photons pairs. 
 This phase manifests as a shift of two-photon
 interference fringes. 
Under certain arrangements, the geometric phase can vary nonlinearly
 and become very sensitive to a change in the polarization
 state. Moreover, it is known that the geometric phase for
 $N$ identically polarized photons is $N$ times larger than that for one
 photon. Thus, the geometric phase for two photons can become two times
 more sensitive to a state change. This high sensitivity to a change in
 the polarization can be exploited for precision measurement of small polarization
 variation. We evaluate the signal-to-noise ratio of
 the measurement scheme using the nonlinear behavior of the geometric phase
 under technical noise and highlight the practical advantages of this
 scheme. 
\end{abstract}

\pacs{03.65.Vf, 42.65.Lm, 42.50.-p}

\maketitle
\section{Introduction}
When a system evolves in such a manner that it returns to its original
state after some time, its wavefunction acquires an additional phase factor, 
which depends solely on the path traced in the ray space. 
The geometric phase was first discovered by Berry in adiabatic, cyclic 
evolution of pure quantum states
\,\cite{berry84:_quant_phase_factor_accom_adiab_chang}.
The geometric phase has been generalized to other state evolutions
including nonadiabatic
evolution\,\cite{aharonov87:_phase_chang_durin_cyclic_quant_evolut,anandan92:_geomet_phase}, 
noncyclic evolution\,\cite{samuel88:_gener_settin_for_berry_phase,morinaga07:_berry_phase_for_noncy_rotat} 
and mixed state evolutions
\,\cite{uhlmann86:_paral_trans_and_quant_holon,sjoqvist00:_geomet_phases_for_mixed_states_in_intery}. 
In optics, Pancharatnam reported the geometric phase in the polarization
state\,\cite{pancharatnam56:_proc}. 
His pioneering work is now widely regarded as being an early precursor of
the geometric
phase\cite{berry87:_adiab_phase_and_panch_phase,samuel88:_gener_settin_for_berry_phase,mukunda93:_quant_kinem_approac_to_geomet_phase}.

There have been many interesting studies on the observation of the geometric
phase\,\cite{tomita86:_obser_of_berry_topol_phase,simon88:_evolv_geomet_phase_and_its,chiao88:_obser_of_topol_phase_by,kwiat91:_obser_of_noncl_berry_phase_for_photon,wagh95:_measur_panch_phase,wagh95:_measur_panch_phase2,loredo09:_measur_of_panch_phase_by,kobayashi11:_obser_of_geomet_phases_in_quant_eraser}. 
Schmitzer \textit{et al.}\,\cite{schmitzer93:_nonlin_of_panch_topol_phase} 
reported that the variation of the geometric phase
exhibits extraordinary nonlinearity associated with post-selection. 
In a certain arrangement, a small change in the pre- or post-selected
state induces a large phase
shift\,\cite{schmitzer93:_nonlin_of_panch_topol_phase,tewari95:_four_arm_sagnac_inter_switc,bhandari97:_polar_of_light_and_topol_phases,hils99:_nonlin_of_pnach_geomet_phase,li99:_exper_obser_of_nonlin_of,tamate09:_geomet_aspec_of_weak_measur,kobayashi11:_obser_of_geomet_phases_in_quant_eraser}. 
Several applications have been proposed that utilize the nonlinear
behavior of the geometric phase including optical
switching\,\cite{tewari95:_four_arm_sagnac_inter_switc,schmitzer91:_optic_switc_based_panch_topol_phase}
and high-precision measurements\,\cite{hils99:_nonlin_of_pnach_geomet_phase,tamate09:_geomet_aspec_of_weak_measur}.

Another important topic is the manifestation of the geometric phase in bipartite and multipartite systems. 
Klyshko\,\cite{klyshko89:_berry_phase_in_multip_exper} showed that the
geometric phase for $N$ identically polarized photons is $N$ times that for one
photon. 
This principle has been observed for two photons in a two-photon
interference experiment utilizing time-correlated photon
pairs\,\cite{brendel95:_geomet_phases_in_two_photon_inter_exper}. 
The effect of entanglement on the geometric phase
has also been discussed in 
\,\cite{sjoeqvist00:_geomet_phase_for_entan_spin_pairs,hessmo00:_quant_phase_for_nonmax_entan_photon,ge05:_geomet_phase_of_entan_spin,williamson07:_compos_geomet_phase_for_multip_entan_states}. 

The aim of the present study is to observe the nonlinear behavior of the geometric
phases of two photons. 
To the best of our knowledge, this is the first observation of the
nonlinear behavior of the two-photon geometric phase. 
In our experiment, time-correlated photon pairs with the same
polarization are incident on a Mach-Zehnder interferometer with polarization elements. 
We can observe the geometric phase of two photons as the phase shift of two-photon
interference fringes using coincidence counting. 
This phase shift is two times larger than that for one photon. 
It lies between $0$ and $4\pi$, i.e., 
the two-photon interference fringe can be shifted by up to two fringe
periods. 
Moreover, the nonlinear behavior suggests that the geometric phase for two
photons is two times more sensitive to a change in the input
polarization than the one-photon case. 
A minute change in the input polarization results in a large shift in the
two-photon interference fringe. 

This high sensitivity to the input polarization can be utilized to
precisely measure small variations. 
We show that the signal-to-noise ratio (SNR) of the measurement scheme
using the geometric phase for multiphoton 
can be improved for a certain type of noise. 
Recently, there has been a related discussion about signal enhancement
with post-selection, the so-called weak measurement amplification, 
in the presence of some noises
\,\cite{aharonov88:_how_resul_of_measur_of,hosten08:_obser_of_spin_hall_effec,dixon09:_ultras_beam_deflec_measur_via,starling09:_optim_signal_to_noise_ratio,feizpour10:_weak_measur_amplif_of_singl_photon_nonlin}.

The remainder of this paper is organized as follows.
In Sec.\,\ref{sec:geom-phas-induc}, 
we briefly review the geometric phase induced by a change in 
the polarization state in a one-photon interferometer and we show that 
the geometric phase can be very sensitive to a change in the
polarization state for a
certain arrangement. Moreover, we show that the $N$-fold geometric phase
for $N$ identically polarized photons can be observed using the same interferometer. 
In Sec.\,\ref{sec:observ-geom-phase}, we introduce the experimental
setup used to observe the geometric phase for two photons
and the results indicating the twofold geometric phase and its
nonlinearity. 
In Sec.\,\ref{sec:discussion-:-snr}, we consider the application of 
the nonlinear behavior of geometric phases to high-precision
measurements. The SNR is evaluated for a practical situation that
includes both shot and technical noise. 
A summary is presented in Sec.\,\ref{sec:summary}.

\section{Geometric phase for $N$ photons and its nonlinearity}
\label{sec:geom-phas-induc}

\subsection{Geometric phases in a one-photon interferometer}
\begin{figure}
\includegraphics[width=7cm]{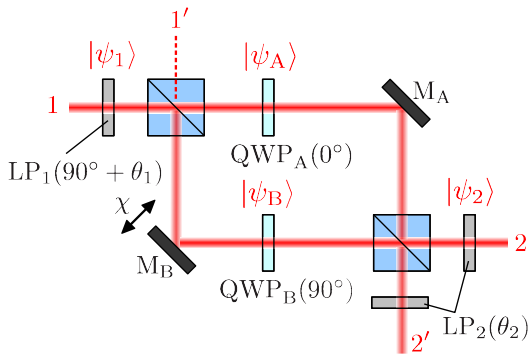}
\caption{(Color online) A Mach-Zehnder interferometer with polarization
 elements used to observe the geometric phase. 
 The interferometer contains two linear polarizers, LP$_1$ and
 LP$_2$, and two quarter-wave plates, QWP$\sub{A}$ and QWP$\sub{B}$.
The angles of the transmission axes of LP$_1$ and LP$_2$ are
 respectively $90\degree+\theta_1$ and $\theta_2$, and those of the fast axes 
of QWP$\sub{A}$ and QWP$\sub{B}$ are
 $0^\circ$ and $90^\circ$. The states $\ket{\psi_1}$, $\ket{\psi_2}$,
 $\ket{\psi\sub{A}}$, and $\ket{\psi\sub{B}}$ are the polarization states
 after LP$_1$, LP$_2$, QWP$\sub{A}$, and QWP$\sub{B}$, respectively. 
}
\label{fig:interferometer}
\end{figure}

We begin by reviewing the geometric phase induced by changing the
polarization state in one-photon interferometer. 
Consider a Mach-Zehnder interferometer with polarization elements as
shown in Fig.\,\ref{fig:interferometer}. In each arm of the
interferometer, the initial polarization state $\ket{\psi_1}$ of an incident photon is
converted into new polarization states $\ket{\psi\sub{A}}$ and
$\ket{\psi\sub{B}}$.
If an additional U(1) phase shift $\chi$ is introduced in one of the
arms, the output intensity $I\sub{m}$ will be
\begin{align}
I\sub{m}
&\propto\left\|\ket{\psi\sub{A}}+\ee^{\ii\chi}\ket{\psi\sub{B}}\right\|^2  \label{eq:1}\\
&= 2\left[1+v\sub{m}\cos(\chi-\phi\sub{m})\right],  \label{eq:2}
\end{align}
where the visibility $v\sub{m}$ and the phase shift $\phi\sub{m}$ are
respectively given by
\begin{align}
v\sub{m}&=\left|\bracketi{\psi\sub{B}}{\psi\sub{A}}\right|,  \label{eq:3}\\
\phi\sub{m}&=\arg\bracketi{\psi\sub{B}}{\psi\sub{A}}.\label{eq:4}
\end{align}
The phase shift $\phi\sub{m}$ expresses the phase difference between the two
different polarization states and is called the relative
phase\,\cite{pancharatnam56:_proc}. 
When $\bracketi{\psi\sub{A}}{\psi\sub{B}}=0$, two states can be perfectly
distinguished and the path followed by the photon is unambiguously
discriminated. 
The interference is then completely destroyed and the
visibility $v\sub{m}$ is reduced to zero.

Next, we consider the phase shift induced by post-selection. 
Post-selection of the polarization state into $\ket{\psi_2}$ causes the output intensity
$I\sub{f}$ to become
\begin{align}
I\sub{f}&\propto\left\|\left(
c\sub{A}+\ee^{\ii\chi}c\sub{B}
\right)
\ket{\psi_2}
\right\|^2  \label{eq:5}\\
&=2p\left[1+v\sub{f}\cos\left(\chi-\phi\sub{f}\right)\right],  \label{eq:6}
\end{align}
where $c\sub{A}=\bracketi{\psi_2}{\psi\sub{A}}$ and
$c\sub{B}=\bracketi{\psi_2}{\psi\sub{B}}$. 
The success probability $p$ of the post-selection, the visibility $v\sub{f}$,
and the phase shift $\phi\sub{f}$ are expressed as
\begin{align}
p&=\frac{1}{2}\left(|c\sub{A}|^2+|c\sub{B}|^2\right),
\label{eq:7}
\\
v\sub{f}&=\frac{2|c\sub{A}c\sub{B}|}
{|c\sub{A}|^2+|c\sub{B}|^2},
\label{eq:8}
\\
\phi\sub{f}&=\arg\bracketi{\psi\sub{B}}{\psi_2}\bracketi{\psi_2}{\psi\sub{A}},
\label{eq:9}
\end{align}
respectively.
Equation (\ref{eq:8}) shows that even when $\ket{\psi\sub{A}}$ is
orthogonal to $\ket{\psi\sub{B}}$, 
the visibility is completely recovered ($v\sub{f}=1$) provided 
$|c\sub{A}|=|c\sub{B}|$. 
In this condition, $\ket{\psi\sub{A}}$ and $\ket{\psi\sub{B}}$ are
projected into the same polarization state $\ket{\psi_2}$ with
the same probability, and it is not possible to determine the photon
paths. This shows that post-selection completely erases the which-path
information and recovers the interference.

\begin{figure}
\includegraphics[width=6cm]{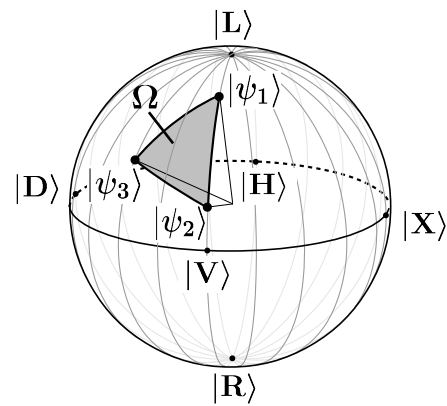}
\caption{Spherical triangle on the Poincar\'{e} sphere formed by three
 polarization states, $\ket{\psi_1}$, $\ket{\psi_2}$, and
 $\ket{\psi_3}$. The geometric phase is proportional to the solid angle
 $\Omega$ of the spherical triangle. The poles correspond
 to the right and left circular polarization states, $\ket{\U{R}}$ and
 $\ket{\U{L}}$, and the equator corresponds to the linear polarization
 states; for example, the horizontal polarization $\ket{\U{H}}$, vertical
 polarization $\ket{\U{V}}$, $45^\circ$ polarization  $\ket{\U{D}}$, 
 and $135^\circ$ polarization $\ket{\U{X}}$.}
\label{fig:pancharatnam_phase}
\end{figure}
The net phase shift induced by the post-selection is calculated as
\begin{align}
\gamma(\psi\sub{A},\psi\sub{B},\psi_2)&\equiv
\phi\sub{f}-\phi\sub{m}  \label{eq:10}\\
&=\arg\bracketi{\psi\sub{A}}{\psi\sub{B}}
\bracketi{\psi\sub{B}}{\psi_2}
\bracketi{\psi_2}{\psi\sub{A}}.
\label{eq:11}
\end{align}
The cyclic form on the right-hand side of Eq.\,(\ref{eq:11}) is gauge invariant (i.e., 
independent of the choice of the phase factor of each state) because the
bra and ket vectors for each state appear pairwise. This phase shift
$\gamma$ is the geometric
phase\,\cite{pancharatnam56:_proc} and 
can be interpreted geometrically on the Poincar\'{e} sphere
as shown in Fig.\,\ref{fig:pancharatnam_phase}. 
The geometric phase $\gamma(\psi_1,\psi_2,\psi_3)$ can be shown to 
be proportional to the solid angle
$\Omega(\psi_1,\psi_2,\psi_3)$ of the spherical triangle connecting the
states $\ket{\psi_1}$, $\ket{\psi_2}$,
and $\ket{\psi_3}$ with geodesic arcs on the Poincar\'{e} sphere
\cite{pancharatnam56:_proc,aravind92:_simpl_proof_of_panch_theor}, i.e., 
\begin{align}
\gamma(\psi_1,\psi_2,\psi_3)
=\frac{1}{2}\Omega(\psi_1, \psi_2, \psi_3).
\label{eq:12}
\end{align}
The sign of the geometric phase is determined by the order of the states.

\subsection{Nonlinearity of geometric phase for one photon}

Here, we consider the nonlinear behavior of the geometric phase for one
photon using the experimental setup shown in
Fig.\,\ref{fig:interferometer}. This is a similar setup to the one used in previous
experiments with a laser light source
\,\cite{li99:_exper_obser_of_nonlin_of,kobayashi11:_obser_of_geomet_phases_in_quant_eraser}. 

The initial polarization state $\ket{\psi_1}$ is prepared by the
linear polarizer LP$_1$:
\begin{align}
\ket{\psi_1}&=
-\sin\theta_1\ket{\text{H}}+\cos\theta_1\ket{\text{V}},  \label{eq:13}
\end{align}
where $\theta_1$ is the angle between the vertical line ($-\pi/2\leq \theta_1\leq\pi/2$) and the
transmission axis of LP$_1$, $\ket{\text{H}}$ is the horizontal
polarization state, and $\ket{\text{V}}$ is the vertical polarization
state. 
The initial state $\ket{\psi_1}$ is changed by 
two quarter-wave plates, QWP$\sub{A}$ and QWP$\sub{B}$, whose 
fast axes are aligned to form angles of $0\degree$ and $90\degree$:
\begin{align}
\ket{\psi\sub{A}}&=-\sin\theta_1\ket{\text{H}}+\ii\cos\theta_1\ket{\text{V}},
\label{eq:14}
\\
\ket{\psi\sub{B}}&=-\ii\sin\theta_1\ket{\text{H}}+\cos\theta_1\ket{\text{V}}.
\label{eq:15}
\end{align} 
Finally, these polarization states are projected into the same state $\ket{\psi_2}$ by the
linear polarizer LP$_2$:
\begin{align}
\ket{\psi_2}=\cos\theta_2\ket{\text{H}}+\sin\theta_2\ket{\text{V}},
\label{eq:16}
\end{align}
where $\theta_2$ is the angle between the horizontal line ($0\leq\theta_2\leq\pi$) and the
transmission axis of LP$_2$. 
Since this setup satisfies
$|\bracketi{\psi_2}{\psi\sub{A}}|=|\bracketi{\psi_2}{\psi\sub{B}}|$, 
the visibility of the interference fringe $v\sub{f}$ becomes unity.

\begin{figure*}
\begin{minipage}{8cm}
\begin{center}
(a)\hspace*{0.5cm}One photon
\includegraphics[width=8cm]{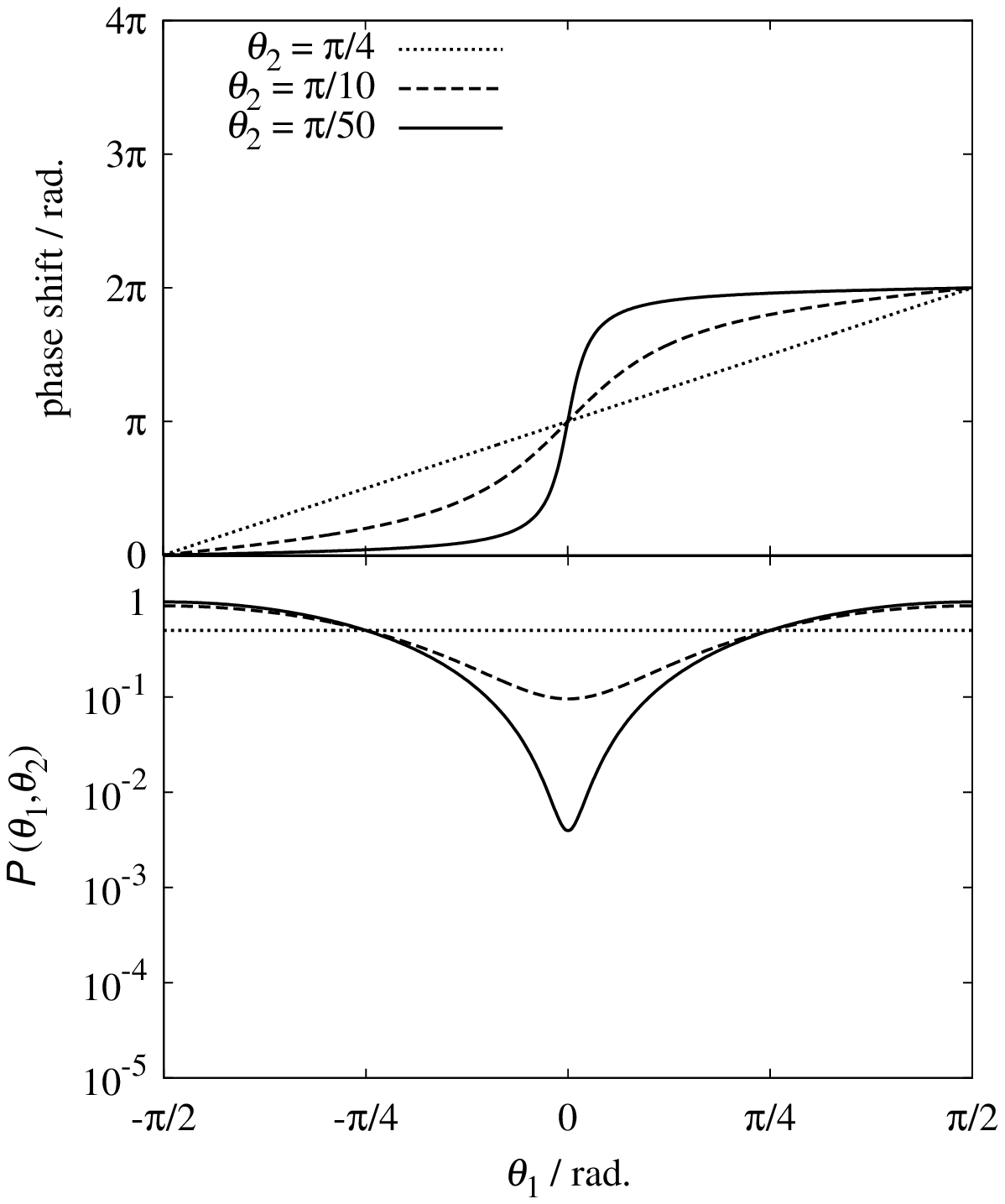}
\end{center}
\end{minipage}
\begin{minipage}{8cm}
\begin{center}
(b)\hspace*{0.5cm}Two photons
\includegraphics[width=8cm]{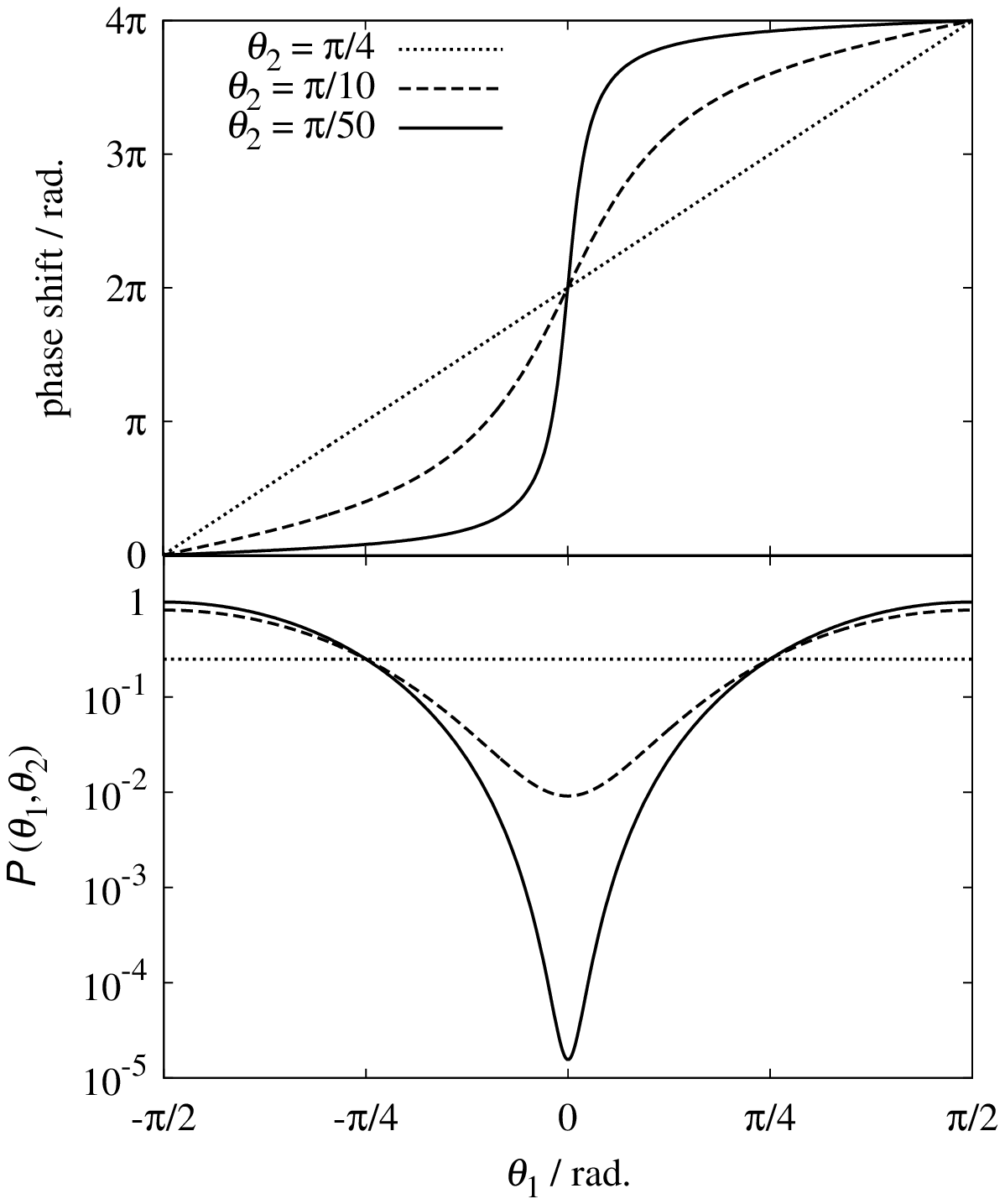}
\end{center}
\end{minipage}
\caption{Geometric phases and success probabilities of post-selection
 for (a) one and (b) two photons. The first row shows the variation of the geometric
 phase and the second row shows the success probability of the
 post-selection with respect to $\theta_1$ for $\theta_2=\pi/50$ (solid line), 
$\theta_2=\pi/10$ (dashed line), and $\theta_2=\pi/4$ (dotted line).}
\label{fig:nonlinear_property}
\end{figure*}

Substituting Eqs.\,(\ref{eq:14}) -- (\ref{eq:16}) into
Eqs.\,(\ref{eq:4}) and (\ref{eq:11}), 
we can obtain the relative phase $\phi\sub{m}$ and
the geometric phase 
$\gamma(\psi\sub{A},\psi\sub{B},\psi_2)$:
\begin{align}
\phi\sub{m}&=
\begin{cases}
\displaystyle\frac{\pi}{2}\hspace*{0.8cm}(\cos 2\theta_1\geq 0),\\
\vspace*{-0.3cm}
\\
\displaystyle-\frac{\pi}{2}\hspace*{0.5cm}(\cos 2\theta_1<0),
\end{cases}
  \label{eq:17}\\
\gamma&=
\begin{cases}
&\displaystyle
\hspace*{-0.1cm}2\tan^{-1}\left(\frac{\tan\theta_1}{\tan\theta_2}\right)\hspace*{0.8cm}
(\cos 2\theta_1\geq 0),\\
\vspace*{-0.3cm}
\\
&\displaystyle
\hspace*{-0.1cm}2\tan^{-1}\left(\frac{\tan\theta_1}{\tan\theta_2}\right)+\pi
\hspace*{0.18cm}(\cos 2\theta_1<0),
\end{cases}
\label{eq:18}
\end{align}
where the range of $\tan^{-1}()$ is $(-\pi/2:\pi/2]$.
The summation of two phases is given by
\begin{align}
\phi\sub{f}=\phi\sub{m}+\gamma
=2\tan^{-1}\left(\frac{\tan\theta_1}{\tan\theta_2}\right)+\frac{\pi}{2}.
\label{eq:19}
\end{align}
The top of Fig.\,\ref{fig:nonlinear_property}(a) shows the variation
of Eq.\,(\ref{eq:18}) with respect to $\theta_1$ for three different values of $\theta_2$. 
It shows that, except for $\theta_2=\pi/4$, the geometric phase is
nonlinear with respect to $\theta_1$. 
The phase shift around $\theta_1=0$ is sensitive to a change in
$\theta_1$ when $\theta_2$ is small.
This nonlinear variation can be observed as a rapid displacement in the
interference fringe when we change $\theta_1$ by rotating LP$_1$ 
\,\cite{schmitzer93:_nonlin_of_panch_topol_phase,tewari95:_four_arm_sagnac_inter_switc,bhandari97:_polar_of_light_and_topol_phases,hils99:_nonlin_of_pnach_geomet_phase,li99:_exper_obser_of_nonlin_of}. 

\begin{figure}
\includegraphics[width=8cm]{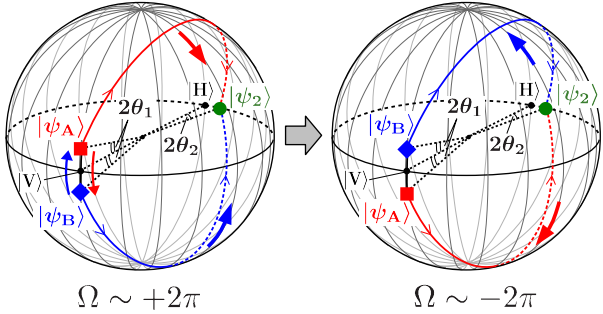}
\caption{(Color online) Geometrical interpretation of the nonlinear variation of the
 geometric phase around 
$(\theta_1,\theta_2)=(0,0)$. 
If $\ket{\psi\sub{A}}$ and $\ket{\psi\sub{B}}$ are
 close to each other across the vertical polarization state
 $\ket{\text{V}}$ on the Poincar\'{e} sphere, 
the area of the spherical triangle will vary rapidly with movement
 of $\ket{\psi\sub{A}}$ and $\ket{\psi\sub{B}}$. 
}
\label{fig:Pancharatnam_on_Poincare1}
\end{figure}

Equation\,(\ref{eq:18}) shows that the nonlinear
behavior of the phase shift originates from the geometric phase $\gamma$ 
and can be understood intuitively in terms of the geometry on the
Poincar\'{e} sphere. 
In the present setup, $\ket{\psi\sub{A}}$ and $\ket{\psi\sub{B}}$ 
given by Eqs.\,(\ref{eq:13}) and (\ref{eq:15}) 
can be depicted at a latitude of $\pm 2\theta_1$ on the
prime meridian and the final state $\ket{\psi_2}$ 
given by Eq.\,(\ref{eq:16}) 
can be depicted on the equator at a longitude of $2\theta_2$ (see
Fig.\,\ref{fig:Pancharatnam_on_Poincare1}). 
When $0<\theta_2<\theta_1\ll 1$,
$\ket{\psi\sub{A}}$ and $\ket{\psi\sub{B}}$ are located near 
the vertical polarization state while $\ket{\psi_2}$ is near the horizontal polarization
state. In this condition, the spherical triangle connecting
$\ket{\psi\sub{A}}$, $\ket{\psi\sub{B}}$, and $\ket{\psi_2}$ almost
degenerates to a great circle. Therefore, the solid angle $\Omega$ is
almost equal to $+2\pi$. 
Now, we consider that $\theta_1$ is changed to exchange the positions of $\ket{\psi\sub{A}}$ and
$\ket{\psi\sub{B}}$ on the Poincar\'{e} sphere.
$\ket{\psi\sub{A}}$ and $\ket{\psi\sub{B}}$ move toward $\ket{V}$ with
decreasing $\theta_1$.
When the distance between $\ket{\psi\sub{A}}$ and $\ket{\psi\sub{B}}$ becomes less than $2\theta_2$, 
the area of the spherical triangle shrinks rapidly. 
After traversing $\ket{\text{V}}$, the area $\Omega$ blows up rapidly and approaches $-2\pi$. 
Thus, the area $\Omega$ changes rapidly from $-2\pi$ to
$2\pi$ around $\theta_1=0$ and the phase shift can vary nonlinearly.

In the nonlinear region of
the phase shift around $\theta_1=0$, 
the success probability of the post-selection drops according
to Eq.\,(\ref{eq:7}):
\begin{align}
p(\theta_1,\theta_2)&=\sin^2\theta_1\cos^2\theta_2+\cos^2\theta_1\sin^2\theta_2.
\label{eq:20}
\end{align} 
The bottom of Fig.\,\ref{fig:nonlinear_property}(a) shows a plot of
Eq.\,(\ref{eq:20}) 
on a logarithmic scale. It implies that a rapid
change in the geometric phase around $\theta_1=0$ can be achieved at the expense of the
output intensity.

\subsection{Geometric phase for $N$ photons}
As shown by Klyshko\,\cite{klyshko89:_berry_phase_in_multip_exper},
$N$ identically polarized photons are expected to acquire $N$ times the
geometric phase for one photon. 
In this section, we theoretically analyze 
the interferometric method for observing the $N$-fold geometric phase.

Assuming that a collection of $N$ photons is incident on the
interferometer (see Fig.\,\ref{fig:interferometer}) and that 
these photons can form path-entangled
states in the interferometer known as $N00N$ states (i.e., all the $N$
photons pass through path A or path B
\,\cite{sanders89:_quant_dynam_of_nonlin_rotat,bollinger96:_optim_frequen_measur_with_maxim_correl_states,boto00:_quant_inter_optic_lithog,edamatsu02:_measur_of_photon_de_brogl,philip04:_de_brogl_wavel_of_non,nagata07:_beatin_stand_quant_limit_with,dowling08:_quant_optic_metrol}), 
the polarization state of the $N$ photons can be expressed as the $N$th tensor product:
\begin{align}
\ket{\Psi_i}\equiv\ket{\psi_i}^{\otimes N},  \label{eq:21}
\end{align}
where $i=(1,2,\text{A},\text{B})$ and $\otimes$ represents the tensor product.

If an additional U(1) phase shift $\chi$ is introduced in one of the
arms, the output intensity $I\sub{m}$ measured by the $N$-photon
coincidence detector will be
\begin{align}
I\sub{m}&\propto
\left\|\ket{\Psi\sub{A}}+\ee^{\ii N\chi}\ket{\Psi\sub{B}}\right\|^2  \label{eq:22}\\
&=2\left[1+V\sub{m}\cos\left(N\chi-\Phi\sub{m}\right)\right],
\label{eq:23}
\end{align}
where the visibility $V\sub{m}$ and the phase shift $\Phi\sub{m}$ are 
respectively given by
\begin{align}
V\sub{m}&=|\bracketi{\psi\sub{B}}{\psi\sub{A}}|^N,  \label{eq:24}\\
\Phi\sub{m}&=N\arg\bracketi{\psi\sub{B}}{\psi\sub{A}}.  \label{eq:25}
\end{align}
Since $N$ photons act as a collective entity in the interferometer,
the phase term in Eq.\,(\ref{eq:23}) is $N$ times that for the one photon case.

After post-selection into the polarization state $\ket{\Psi_2}$, the
output intensity $I\sub{f}$ is given by
\begin{align}
I\sub{f}&\propto
\left\|
\left(\bracketi{\Psi_2}{\Psi\sub{A}}
+\ee^{\ii N\chi}\bracketi{\Psi_2}{\Psi\sub{B}}\right)\ket{\Psi_2}
\right\|^2  \label{eq:26}\\
&=2P\left[1+V\sub{f}\cos\left(N\chi-\Phi\sub{f}\right)\right],  \label{eq:27}
\end{align}
where the success probability for post-selection $P$, the visibility
$V\sub{f}$, and the phase shift $\Phi\sub{f}$ are respectively expressed by
\begin{align}
P&=\frac{1}{2}\left(|c\sub{A}|^{2N}+|c\sub{B}|^{2N}\right),  \label{eq:28}\\
V\sub{f}
&=\frac{2|c\sub{A}c\sub{B}|^N}
{|c\sub{A}|^{2N}+|c\sub{B}|^{2N}}  \label{eq:29}\\
\Phi\sub{f}&=N\arg\bracketi{\psi\sub{B}}{\psi_2}\bracketi{\psi_2}{\psi\sub{A}}.  \label{eq:30}
\end{align}
Substituting Eqs.\,(\ref{eq:14}) -- (\ref{eq:16}) into 
Eqs.\,(\ref{eq:28}) -- (\ref{eq:30}) we can obtain
\begin{align}
P&=\left(\sin^2\theta_1\cos^2\theta_2
+\cos^2\theta_1\sin^2\theta_2\right)^N,  \label{eq:31}\\
V\sub{f}&=1,  \label{eq:32}\\
\Phi\sub{f}&=
2N\tan^{-1}
\left(\frac{\tan\theta_1}{\tan\theta_2}\right)+\frac{N\pi}{2}.  \label{eq:33}
\end{align}
Figure\,\ref{fig:nonlinear_property}(b) shows the variation of $P$ and
$\Phi\sub{f}$ with respect to $\theta_1$ for $N=2$. 

Since the slope of the phase shift for $N$ photons around
$\theta_1=0$ is $N$ times steeper than that for one 
photon, we can obtain an $N$-fold enhancement in the variation of
$\theta_1$ [see the top of Fig.\,\ref{fig:nonlinear_property}].
However, the success probability of post-selection $P$ decreases as the
$N$th power of the one-photon success probability $p$ because
$P$ corresponds to the probability that all $N$ photons are 
successfully post-selected into state $\ket{\psi_2}$ [see the bottom of
Fig.\,\ref{fig:nonlinear_property}]. 

\section{Observation of geometric phase for two photons using
 photon pairs}
\label{sec:observ-geom-phase}
\begin{figure}
\begin{center}
\includegraphics[width=8cm]{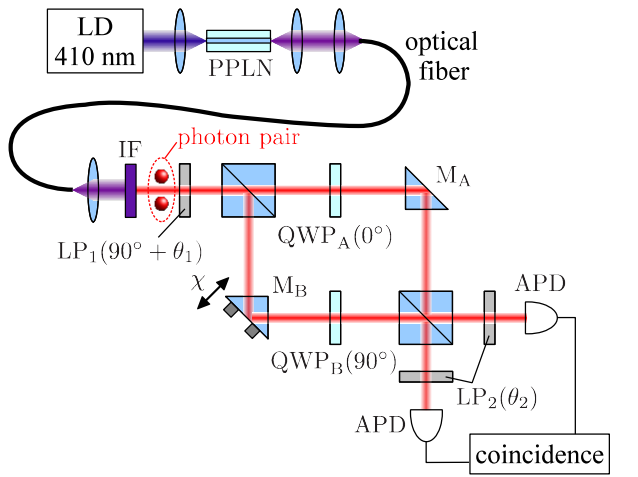}
\caption{(Color online) Experimental setup for observing the geometric
 phase for two photons. IF is an interference filter and M$\sub{A}$ and
 M$\sub{B}$ are total reflection mirrors.}
\label{fig:setup}
\end{center}
\end{figure}

\subsection{Two-photon interference in Mach-Zehnder interferometer}
We consider a two-photon state input at port 1 and a vacuum state
input at port $1^\prime$ of a symmetric Mach-Zehnder interferometer. 
The first beam splitter splits the incident photons 
into two paths A and B: 
\begin{align}
&\ket{2}_{1}\ket{0}_{1^\prime}\nonumber\\
\rightarrow&
\ket{2}\sub{A}\ket{0}\sub{B}+\ee^{\ii 2\chi}\ket{0}\sub{A}\ket{2}\sub{B}
+2\ee^{\ii\chi}\ket{1}\sub{A}\ket{1}\sub{B},  \label{eq:34}
\end{align}
where the photon number state with $N$ photons in path (or port) $p$ is written as
 $\ket{N}_p$ and $\chi$ is the additional phase shift.
When we operate coincidence counting between output port $2$ and
$2^\prime$, 
the $\ket{1}\sub{A}\ket{1}\sub{B}$ term in Eq.\,(\ref{eq:34}) is missing
due to the complete destructive two-photon
interference\,\cite{hong87:_measur_of_subpic_time_inter}, producing a 
two-photon path-entangled state to be detected. 

The above consideration is valid even if the interferometer contains
polarization elements as shown in Fig.\,\ref{fig:interferometer} because
the polarization states of two photons are eventually projected into the same
states with the same probability
at output ports $2$ and $2^\prime$. 
Hence, we can observe the nonlinear behavior of the phase shift of 
two photons using coincidence counting between output ports $2$ and $2^\prime$.

\subsection{Experimental setup}
\begin{figure}
\begin{center}
\includegraphics[width=9cm]{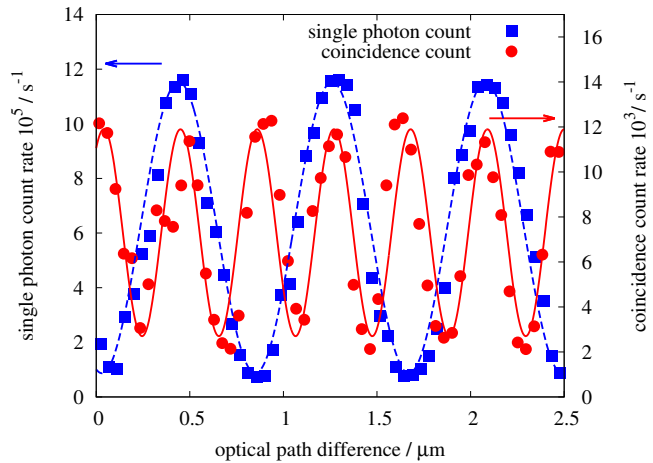}
\caption{(Color online) Measured one- and two-photon interference fringes. 
The blue squares (\textcolor{blue}{\tiny$\blacksquare$}) and red circles
 (\textcolor{red}{$\bullet$}) indicate the one-photon interference measured 
by single-photon counting and the two-photon interference measured by 
coincidence counting, respectively. 
The solid curves show the theoretical interferences. }
\label{fig:fringe}
\end{center}
\end{figure}
\begin{figure*}
\begin{minipage}{7cm}
\begin{center}
(a)\hspace*{0.5cm}One photon\\
\hspace*{-1cm}\includegraphics[width=8.5cm]{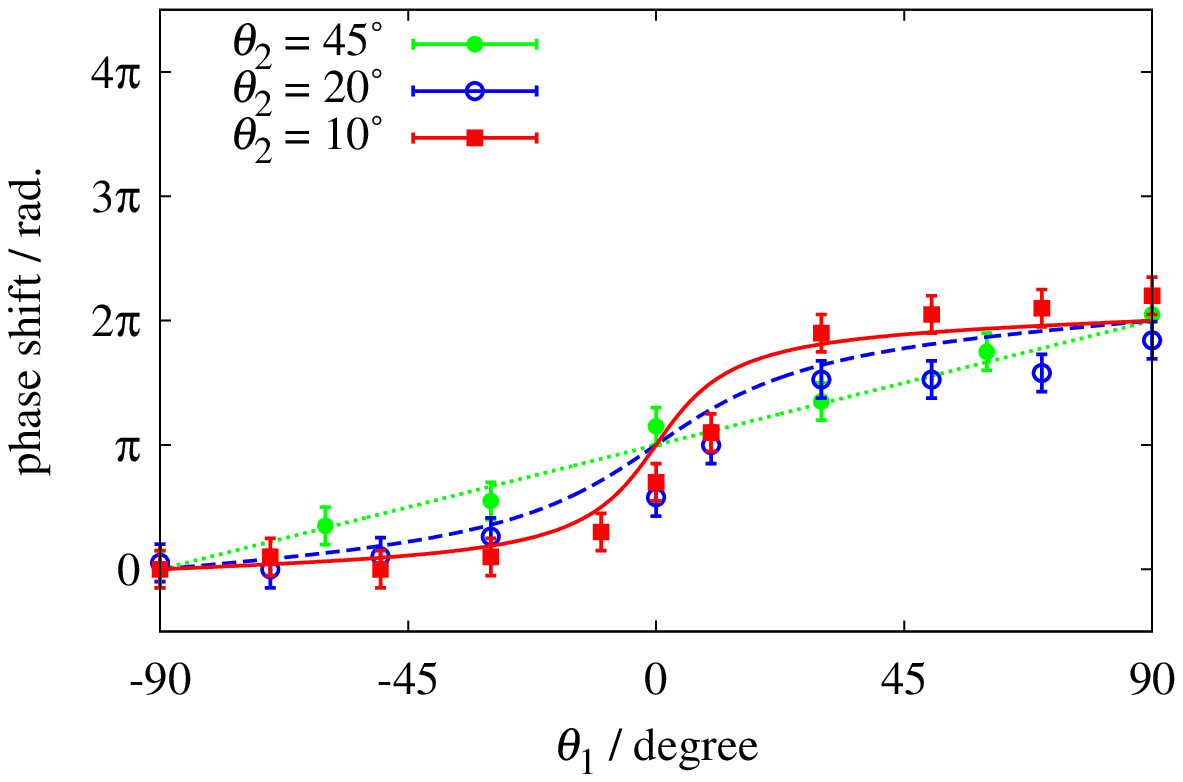}
\end{center}
\end{minipage}
\begin{minipage}{8cm}
\begin{center}
(b)\hspace*{0.5cm}Two photons
\includegraphics[width=8.5cm]{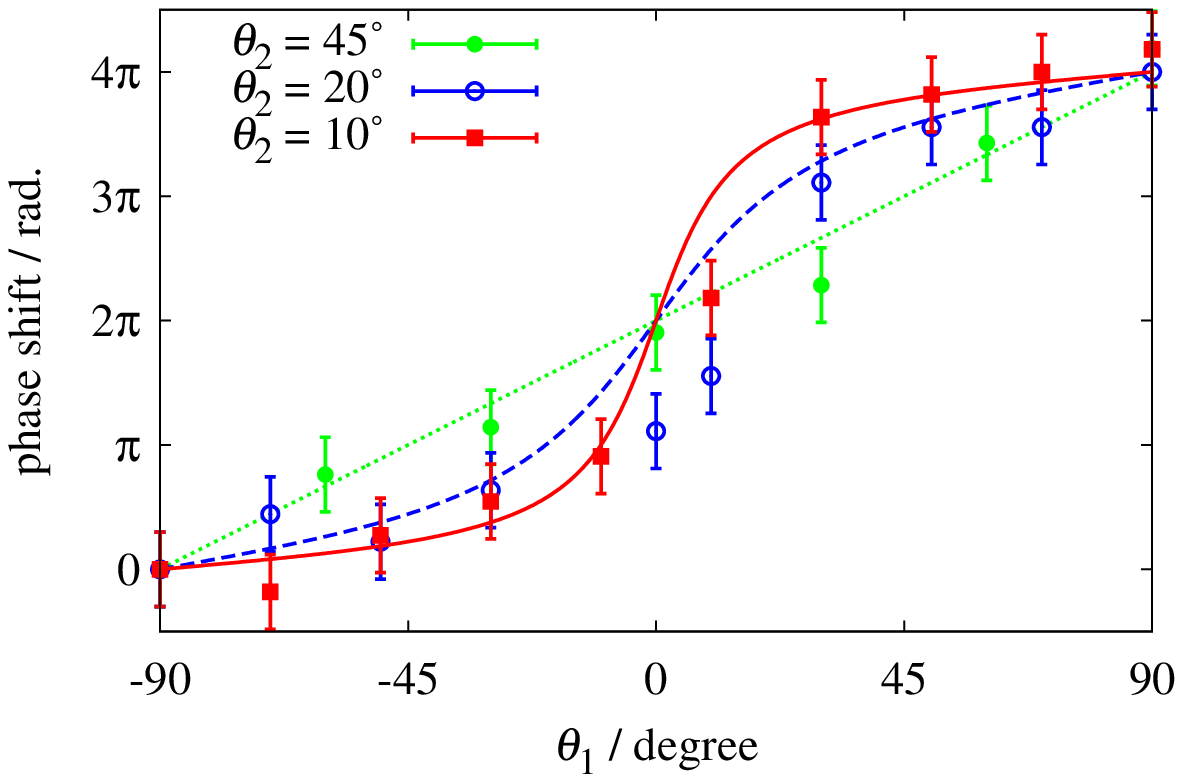}
\end{center}
\end{minipage}
\caption{(Color online) Experimental results for geometric phases of
 (a) one and (b) two photons with respect to
 $\theta_1$ for three different values of $\theta_2$.}
\label{fig:phase_shift}
\end{figure*}
Fig.\,\ref{fig:setup} shows a schematic representation 
of the experimental setup. 
Photon pairs are generated in the periodically poled LiNbO$_3$
(PPLN) waveguide via degenerate type-I parametric down-conversion
of a 410-nm blue light from a laser diode (LD). 
The temperature of PPLN is calibrated by a temperature controller 
to satisfy the phase matching condition.
The center wavelength of the photon pairs is $820$ $\U{nm}$ and its spectral
bandwidth is restricted to $20$ $\U{nm}$ by the interference filter IF. 
After beam shaping by the single-mode optical fiber, photon pairs traverse
the Mach-Zehnder interferometer with polarization elements. 
The phase difference between two arms is varied continuously 
by shifting the total reflection mirror M$\sub{B}$ using a piezoelectric 
translation stage. The outputs of the interferometer are coupled to a pair of single-photon
counting modules (Perkin Elmer, SPCM-AQR-14).
Individual photon counts and coincidence counts are recorded using
field-programmable gate array (FPGA) electronics connected to a personal
computer\,\cite{branning09:_low_cost_coinc_count_elect}.

\subsection{Observation of geometric phases for one and two photons}

Figure\,\ref{fig:fringe} shows the one- and two-photon interference
fringes obtained for photon pairs. 
The solid line is a fit by a sinusoidal function. 
The vertical axis shows the single-photon and the coincidence
count rates. 
This figure shows that the two-photon interference fringe has a period given by the
wavelength of the pump light and an average visibility of $63\%$. 

Figure\,\ref{fig:phase_shift} shows the phase shifts of one-photon and
two-photon interference fringes with respect to $\theta_1$ 
for $\theta_2=45\degree$ (filled green circles), 
$20\degree$ (open blue circles), and $10\degree$ (filled red squares). 
The origin of the vertical axis is determined by the position of 
fringes when $\theta_1=-90\degree$ and the value of the vertical axis
shows the displacement of fringes normalized by one period of the fringes. 
The solid line in Fig.\,\ref{fig:phase_shift} indicates the theoretical curve
calculated from Eq.\,(\ref{eq:33}). 
For both one- and two-photon interference, 
the gradient of the variation of the phase shift around
$\theta_1=0\degree$ increases with decreasing $\theta_2$. 
This implies that the variation in the phase shift becomes more
sensitive to a variation in $\theta_1$. 
Moreover, the phase shift for two photons is two times larger than that
for one photon. Thus, the gradient of the phase shift for two photons around
$\theta_1=0\degree$ also becomes two times steeper than that for one photon.

\section{Discussion : SNR of measurement scheme using nonlinear behavior of
 geometric phase}
\label{sec:discussion-:-snr}
In what follows, we consider the measurement of a small polarizer angle
$\theta_1$ ($|\theta_1|\ll 1$) through the phase shift of the interference
fringes. 
As shown in the previous section, the geometric phase becomes sensitive
to a variation in $\theta_1$ around $\theta_1=0$ when $\theta_2$ is
small. 
Moreover, the geometric phase for $N$ photons will be $N$ times more
sensitive to a variation in $\theta_1$. 
Utilizing this nonlinear behavior, we can measure the small angle $\theta_1$
from the large phase shift of the ($N$-photon) interference fringe. 
However, the small success probability of post-selection around
$\theta_1=0$ might cancel out the advantage of the large phase shift. 

In this section, we calculate the SNR of this measurement using the geometric
phase for $N$ photons to evaluate its advantages and disadvantages. 
Interferometric phase measurement is subject to various 
noises. In the ideal situation, the shot noise is dominant, whereas in 
most experiments, the SNR is limited by technical noises such as 
excessive fluctuations in the light sources. 
Thus, we calculate the SNR 
for a case with technical noise in addition to the shot noise. 
We show that the nonlinearity of the geometric phase does not
improve the SNR in the shot noise limit. 
However, under certain technical noises, the large phase shift due
to the nonlinearity of the geometric phase can be of practical advantage.

\subsection{SNR of direct measurement}
\begin{figure}
\includegraphics[width=7.5cm]{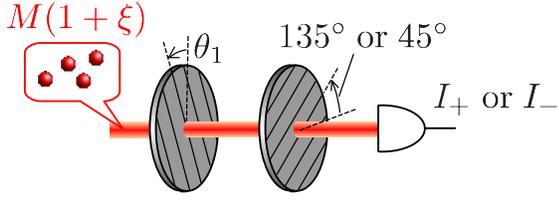}
\caption{(Color online) Direct measurement of $\theta_1$. $I_+$ and $I_-$ show the
 single photon counts for $135\degree$ and $45\degree$ polarization,
 respectively.}
\label{fig:direct_estimation}
\end{figure}
First, we evaluate the SNR of the direct measurement of $\theta_1$
without utilizing the geometric phase as shown in
Fig.\,\ref{fig:direct_estimation}. 
We consider a sequential measurement of 
single-photon counts integrated over the
interval $\tau$ for $45\degree$ and $135\degree$ polarizations expressed as 
\begin{align}
\bracket{I_+(\tau)}&=\eta\tau M|\bracketi{\text{X}}{\psi_1}|^2
=\frac{\eta\tau M}{2}\left(
1+\sin 2\theta_1
\right),  \label{eq:35}\\
\bracket{I_-(\tau)}&=\eta\tau M|\bracketi{\text{D}}{\psi_1}|^2
=\frac{\eta\tau M}{2}\left(
1-\sin 2\theta_1
\right),  \label{eq:36}
\end{align}
where $\bracket{\hspace*{0.1cm}\cdot\hspace*{0.1cm}}$ shows the ensemble
mean value, $M$ is the number of incident photons per unit time, and
$\eta$ is the detection efficiency. 
The angle $\theta_1$ can be measured 
from the ratio of the difference to the sum between $\bracket{I_+}$ and
$\bracket{I_-}$:
\begin{align}
\bracket{n(\theta_1)}&\equiv\frac{\bracket{I_+(\tau)}-\bracket{I_-(\tau)}}
{\bracket{I_+(\tau)}+\bracket{I_-(\tau)}}  \label{eq:37}\\
&\simeq 2\theta_1,  \label{eq:38}
\end{align}
where we assume that the absolute value of $\theta_1$ is small
($|\theta_1|\ll 1$). 
The sum contains information about the total number of successful
measurements, and the difference contains phase information
as a function of $\theta_1$. 
From Eq.\,(\ref{eq:38}), the mean value of experimentally obtained counts $\bracket{n}$ 
is twice the true value $\theta_1$.

Here, we consider the technical noise due to the fluctuation in the
number of incident photons (i.e., $M\rightarrow M(1+\xi(t))$) in
addition to the shot noise. 
Since we assume that two outputs $I_+$ and $I_-$ are measured
sequentially, the technical noises in the two outputs are uncorrelated
with each other and there remains technical noise in the difference
of the two outputs. 
Under this condition, the variance of the total noise around $\theta_1=0$
is calculated from Eq.\,(\ref{eq:67}) as
\begin{align}
\bracket{\delta n(\tau)^2}
=\frac{1}{\tau}
\left(\frac{1}{\eta M}+\frac{\bar{\xi}^2}{2}\right),  \label{eq:39}
\end{align}
where $\bar{\xi}^2$ is the power spectrum of the intensity
fluctuation. 
The first and second terms of Eq.\,(\ref{eq:39}) 
are respectively attributed to the shot noise and the
technical noise. 
The shot noise is dependent on $M$, whereas 
the technical noise is independent of it. Thus, we cannot
reduce the technical noise by increasing the input beam intensity. 

From Eqs.\,(\ref{eq:37}) and (\ref{eq:39}), the SNR of the direct
measurement is obtained as
\begin{align}
\mathcal{R}&=
\frac{\bracket{n(\theta_1)}}{\sqrt{\bracket{\delta n(\tau)^2}}}  \label{eq:40}\\
&\simeq\frac{2\sqrt{\tau}\theta_1}{\sqrt{(\eta M)^{-1}+\bar{\xi}^2/2}}.  \label{eq:41}
\end{align}
The above equation shows that the SNR is proportional to $\sqrt{M}$ in the
region where the shot noise is dominant (i.e., $\eta
M\ll\bar{\xi}^{-2}$) 
whereas the SNR is constant with respect to $M$ in the region where the
technical noise is dominant (i.e., $\eta M\gg\bar{\xi}^{-2}$). 

\subsection{SNR of measurement utilizing the nonlinearity of the geometric phase}
We now derive the SNR of the measurement using the nonlinear behavior
of the geometric phases.
We consider the sequential measurement of 
the two different outputs, $I_+$ and $I_-$,
corresponding to fringes of $N$-photon interference
that are out of phase with each other:
\begin{align}
\bracket{I_+(\tau)}&=\frac{\eta\nu\tau P(\theta_1,\theta_2)}{2}
\left\{1+V\sub{f}\cos\left[\Phi(\theta_1,\theta_2)\right]\right\},
\label{eq:42}\\
\bracket{I_-(\tau)}&=\frac{\eta\nu\tau P(\theta_1,\theta_2)}{2}
\left\{1-V\sub{f}\cos\left[\Phi(\theta_1,\theta_2)\right]\right\},
\label{eq:43}
\end{align}
where $\Phi(\theta_1,\theta_2)=N\chi-\Phi\sub{f}(\theta_1,\theta_2)$, 
$\eta$ is the detection efficiency of $N$-photon coincidence, 
$\nu$ is the incident $N$-photon flux per unit time, 
and $\tau$ is the integrated time for $N$-photon coincidence counting. 
The total photon number per unit time is $M=\nu N$. 

The angle $\theta_1$ can be measured from $\bracket{I_+}$ and
$\bracket{I_-}$ as
\begin{align}
\bracket{n(\theta_1,\theta_2)}&=\frac{\bracket{I_+(\tau)}-\bracket{I_-(\tau)}}
{\bracket{I_+(\tau)}+\bracket{I_-(\tau)}}  \label{eq:44}\\
&=V\sub{f}\cos\left[\Phi(\theta_1,\theta_2)\right].  \label{eq:45}
\end{align}
To measure a small value of $\theta_1$, the offset phase $\chi$ is set to
satisfy $\bracket{n(\theta_1=0,\theta_2)}=0$; i.e., 
\begin{align}
\chi=\frac{N+1}{2N}\pi.  \label{eq:46}
\end{align}
In this condition, $\bracket{n}$ is calculated as
\begin{align}
\bracket{n(\theta_1,\theta_2)}&=
V\sub{f}
\sin\left[
2N\tan^{-1}\left(
\frac{\tan\theta_1}{\tan\theta_2}
\right)
\right].  \label{eq:47}
\end{align}
For a sufficiently small value of $\theta_1$ satisfying
\begin{align}
\frac{|\theta_1|}{\tan\theta_2}\ll\tan\frac{1}{2N},  \label{eq:48}
\end{align}
$\bracket{n}$ is found to be
\begin{align}
\bracket{n(\theta_1,\theta_2)}&
\simeq V\sub{f}\hspace*{0.05cm}\theta_1
\fracpd{\Phi\sub{f}}{\theta_1}\Bigr|_{\theta_1=0}  \label{eq:49}\\
&=
\frac{2NV\sub{f}}{\tan\theta_2}\theta_1.  \label{eq:50}
\end{align}
Comparing Eq.\,(\ref{eq:50}) with Eq.\,(\ref{eq:38}), we find that the
experimentally obtained value is enhanced by the gradient of the geometric
phase $NV\sub{f}/\tan\theta_2$. 

In the same manner as Eq.\,(\ref{eq:39}), we introduce fluctuation in
the incident $N$-photon flux $\nu$; i.e.,
$\nu\rightarrow\nu(1+\xi(t))$. 
This type of noise may be introduced via intensity
fluctuations of the pump beam driving $N$-photon generation.
The variance in the total noise around $\theta_1=0$ is calculated as
\begin{align}
\bracket{\delta n(\tau)^2}=\frac{1}{\tau}\left(
\frac{1}{\eta\nu P(0,\theta_2)}+\frac{\bar{\xi}^2}{2}
\right),  \label{eq:51}
\end{align}
where the technical noises in $I_+$ and $I_-$ are assumed to be
uncorrelated with each other. 
Comparing the above equation with Eq.\,(\ref{eq:39}), we can see that the technical noise is
unchanged, whereas the shot noise is increased by a factor of
$1/P(0,\theta_2)\simeq (1/\theta_2)^{2N}$ because the number of successful
measurements is reduced due to the small success probability of the post-selection.

\begin{figure}
\begin{center}
\includegraphics[width=8.5cm]{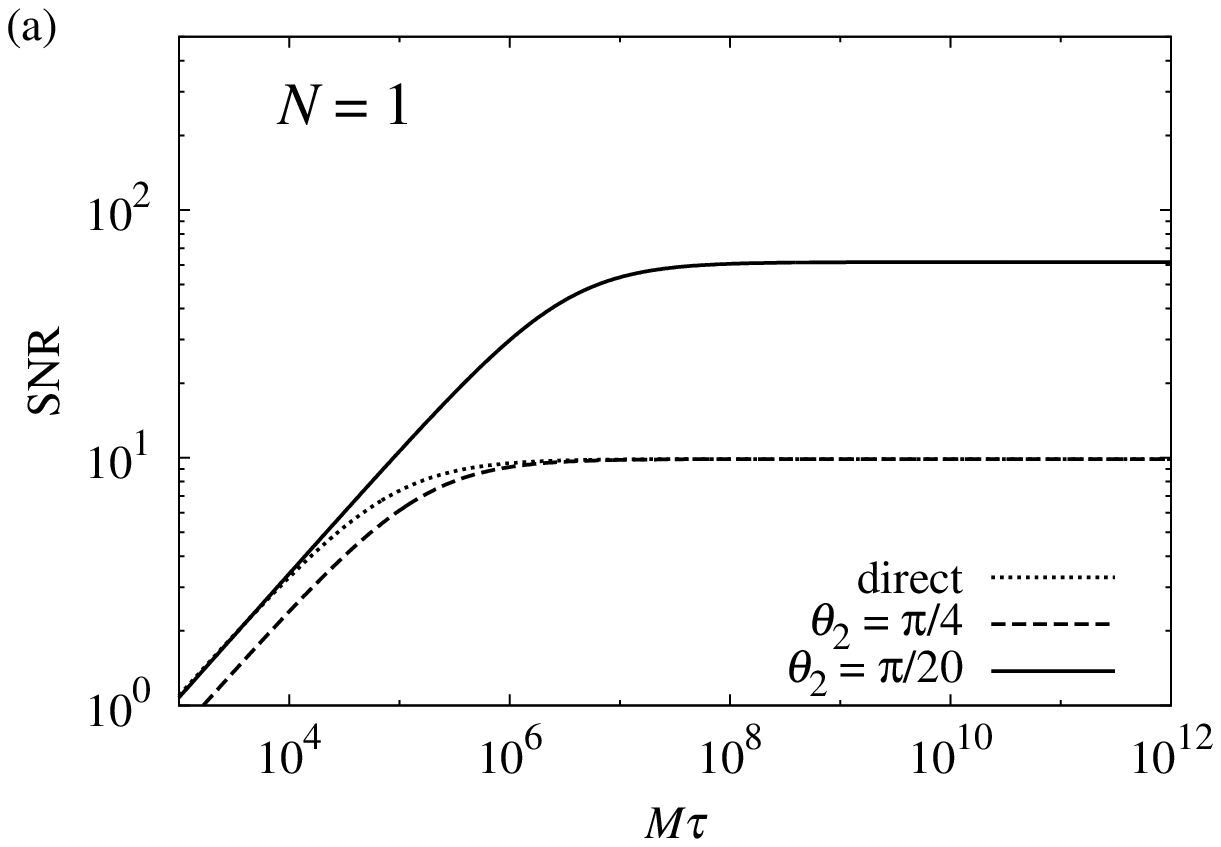}\\
\includegraphics[width=8.5cm]{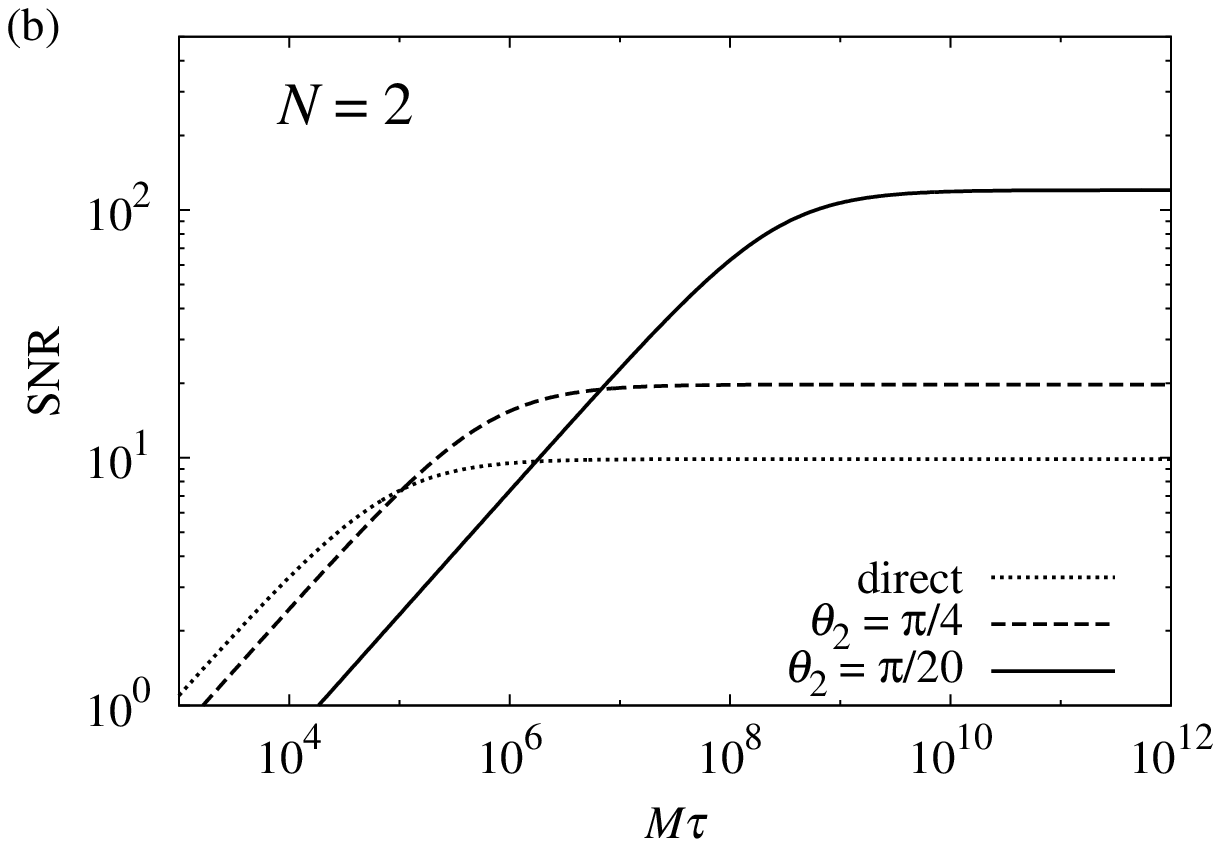}
\end{center}
\caption{The SNR of $\theta_1$ measurement with respect to the total number of
 photons for two different values of $\theta_2$. (a) One- and (b)
 two-photon cases. We have taken $V\sub{f}=1$, $\eta=1.0$, 
$\bar{\xi}^2/\tau=2.5\times 10^{-5}$, and $\theta_1=\pi/180$} 
\label{fig:SN_1+2}
\end{figure}

\begin{figure}
\begin{center}
\includegraphics[width=8.5cm]{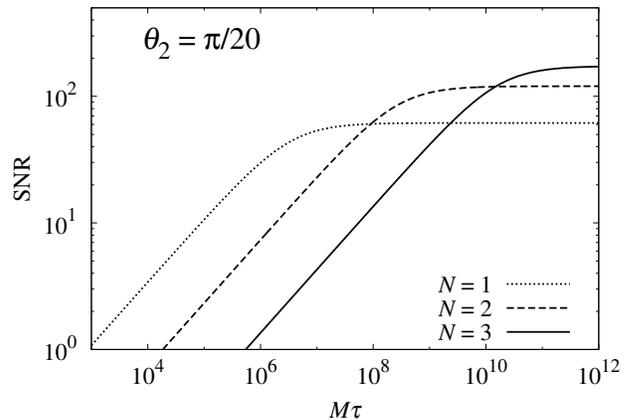}
\end{center}
\caption{The SNR of $\theta_1$ measurement with respect to the total number of
 photons for three different values of $N$. $\theta_2$ is fixed at
 $\pi/20$. We have taken $V\sub{f}=1$, $\eta=1.0$,
 $\bar{\xi}^2/\tau=2.5\times 10^{-5}$, and $\theta_1=\pi/180$.}
\label{fig:SN_N_dependence}
\end{figure}

The SNR is calculated from Eqs.\,(\ref{eq:50}) and (\ref{eq:51}) as
\begin{align}
\mathcal{R}&=\frac{\bracket{n(\theta_1,\theta_2)}}
{\sqrt{\bracket{\delta n(\tau)^2}}}  \label{eq:52}\\
&=\frac{2NV\sub{f}\hspace*{0.05cm}\theta_1}{\tan\theta_2}
\sqrt{
\frac{\tau}{\{N/\eta MP(0,\theta_2)\}+\bar{\xi}^2/2}
}.  \label{eq:53}
\end{align}
As the rate of $M$ increases, the SNR scales as 
$\sqrt{M}$ in the region where the shot noise is
dominant ($\eta\nu P\ll\bar{\xi}^{-2}$). On the other hand, the SNR is saturated 
in the region where the technical noise is dominant ($\eta\nu
P\gg\bar{\xi}^{-2}$). In the latter region, the SNR is enhanced by a
factor of $NV\sub{f}/\tan\theta_2$ compared to Eq.\,(\ref{eq:41}).

\subsection{Comparison of measurement using geometric phase and
  direct measurement}
Figure\,\ref{fig:SN_1+2}
shows a plot of Eqs.\,(\ref{eq:41}) and (\ref{eq:53}) with respect
to the total number photons per unit time, $M=\nu N$, 
for $N=1$ and $N=2$. The dotted line indicates the SNR of direct
measurement. The dashed and solid lines
show the SNR using the geometric phase for $\theta_2=\pi/4$ and
$\pi/20$, respectively. 
For $N=1$ and $N=2$, the SNR using the geometric phase
is smaller than that of direct measurement in the region where the shot
noise is dominant, whereas in the region where the technical noise is
dominant, the SNR is improved by enhancement due to the geometric
phase. In the latter region, the SNR increases with decreasing
$\theta_2$. 
Moreover, the SNR for $N=2$ is two times larger than
that for $N=1$. 

Figure\,\ref{fig:SN_N_dependence} shows the SNR with respect to
$M$ for $N=1$ (dotted line), $2$ (dashed line), and $3$ (solid
line). The maximum SNR for the same $\theta_2$ is proportional to $N$.
Thus, whenever a sufficiently intense beam satisfying $\eta\nu
P\gg\xi^2/2$ 
is used, the SNR can be improved via the
nonlinearity of the $N$-photon geometric phase.

\section{Summary}
\label{sec:summary}
We have shown that the $N$-fold geometric phase manifests in the
$N$-photon interference fringe. 
In our experiment using photon pairs, we obtained the geometric
phase for two photons, and confirmed that it is two times larger
than that for one photon. 
The gradient of the phase shift for two photons is 
also two times steeper than that for one photon. 
We compared the SNRs for a direct measurement and for a 
measurement using the nonlinear behavior of the geometric phase for $N$
photons. 
We demonstrated that the measurement using the nonlinear behavior of the
geometric phase has practical advantages under certain types of 
technical noise. Moreover, it has been shown that 
the SNR using $N$ photons can be $N$ times
larger than that for the one-photon case.

\begin{acknowledgments}
This research is supported by the Global COE Program ``Photonics and
 Electronics Science and Engineering'' at Kyoto University.
\end{acknowledgments}

\begin{appendix}
\section{Mean value and fluctuation of photon counting}
In what follows, $\bracket{\hspace*{0.1cm}\cdot\hspace*{0.1cm}}$ and 
$\delta\hspace*{0.1cm}\cdot$ respectively show the ensemble mean
value and the zero-mean fluctuation. 

We represent the photon counting rate at given time $t$ as
\begin{align}
i(t)=\bracket{i}+\delta i(t).  \label{eq:54}
\end{align}
The fluctuation $\delta i(t)$ contains the technical noise
$\xi(t)$ in addition to the shot noise $\epsilon(t)$. 
The correlation function of the fluctuation is expressed as
\begin{align}
\bracket{\delta i(t)\delta i(t^\prime)}
&=\bracket{\epsilon(t)\epsilon(t^\prime)}+\bracket{\xi(t)\xi(t^\prime)},  \label{eq:55}
\end{align}
where the shot noise and technical noise are assumed to be uncorrelated
with each other. 
Assume that photon detection is a random process and that
$\xi(t)$ can be modeled as white noise:
\begin{align}
\bracket{\epsilon(t)\epsilon(t^\prime)}
&=\bracket{i}\delta(t-t^\prime),  \label{eq:56}\\
\bracket{\xi(t)\xi(t^\prime)}
&=\bar{\xi}^2\delta(t-t^\prime),  \label{eq:57}
\end{align}
where $\delta(t)$ is the Dirac delta function and $\bar{\xi}^2$ is 
the power spectrum of the technical noise.

Integrating the coincidence count over time $\tau$, we obtain
\begin{align}
I(t,\tau)=\bracket{I(\tau)}+\delta I(t,\tau),  \label{eq:58}
\end{align}
where its ensemble mean value and fluctuation are
\begin{align}
\bracket{I(\tau)}&=\bracket{i}\tau,  \label{eq:59}\\
\delta I(t,\tau)&=\int_{t}^{t+\tau}\delta i(t^\prime)\dd t^\prime.  \label{eq:60}
\end{align}
The variance of the noise in $I(t,\tau)$ is given by the ensemble
mean value of $\delta I(t,\tau)^2$:
\begin{align}
\bracket{\delta I(\tau)^2}&=\int_0^\tau\int_0^\tau \bracket{\delta i(t^\prime)
\delta i(t^{\prime\prime})}\dd t^\prime\dd t^{\prime\prime}
 \label{eq:61} \\
&=\left(
\bracket{i}+\bar{\xi}^2
\right)\tau.  \label{eq:62}
\end{align}

To measure a certain parameter, consider the ratio of
the difference to the sum between two outputs $I_1(t_1,\tau)$ and $I_2(t_2,\tau)$:
\begin{align}
n(t_1,t_2,\tau)&\equiv
\frac{I_1(t_1,\tau)-I_2(t_2,\tau)}
{I_1(t_1,\tau)+I_2(t_2,\tau)}  \label{eq:63}\\
&\simeq\bracket{n(\tau)}+\delta n(t_1,t_2,\tau),   \label{eq:64}
\end{align}
with
\begin{align}
\bracket{n(\tau)}&=\frac{\bracket{I_1(\tau)}-\bracket{I_2(\tau)}}
{\bracket{I_1(\tau)}+\bracket{I_2(\tau)}},  \label{eq:65}\\
\delta n(t_1,t_2,\tau)&=
\frac{\delta I_1(t_1,\tau)-\delta I_2(t_2,\tau)}
{\bracket{I_1(\tau)}+\bracket{I_2(\tau)}},  \label{eq:66}
\end{align}
where we assume that $\delta I_1,\delta I_2\ll \bracket{I_1(\tau)}+\bracket{I_2(\tau)}$. 
The variance of the noise in $n(t_1,t_2,\tau)$ is given by
\begin{align}
&\bracket{\delta n(\tau)^2}
=\frac{
\bracket{\left[\delta I_1(t_1,\tau)-\delta I_2(t_2,\tau)\right]^2}}
{\left[\bracket{I_1(\tau)}+\bracket{I_2(\tau)}\right]^2}.  \label{eq:67}
\end{align}

\end{appendix}


\begin{thebibliography}{46}
\expandafter\ifx\csname natexlab\endcsname\relax\def\natexlab#1{#1}\fi
\expandafter\ifx\csname bibnamefont\endcsname\relax
  \def\bibnamefont#1{#1}\fi
\expandafter\ifx\csname bibfnamefont\endcsname\relax
  \def\bibfnamefont#1{#1}\fi
\expandafter\ifx\csname citenamefont\endcsname\relax
  \def\citenamefont#1{#1}\fi
\expandafter\ifx\csname url\endcsname\relax
  \def\url#1{\texttt{#1}}\fi
\expandafter\ifx\csname urlprefix\endcsname\relax\def\urlprefix{URL }\fi
\providecommand{\bibinfo}[2]{#2}
\providecommand{\eprint}[2][]{\url{#2}}

\bibitem[{\citenamefont{Berry}(1984)}]{berry84:_quant_phase_factor_accom_adiab%
_chang}
\bibinfo{author}{\bibfnamefont{M.~V.} \bibnamefont{Berry}},
  \bibinfo{journal}{Proc. R. Soc. London A} \textbf{\bibinfo{volume}{392}},
  \bibinfo{pages}{45} (\bibinfo{year}{1984}).

\bibitem[{\citenamefont{Aharonov and
  Anandan}(1987)}]{aharonov87:_phase_chang_durin_cyclic_quant_evolut}
\bibinfo{author}{\bibfnamefont{Y.}~\bibnamefont{Aharonov}} \bibnamefont{and}
  \bibinfo{author}{\bibfnamefont{J.}~\bibnamefont{Anandan}},
  \bibinfo{journal}{Phys. Rev. Lett.} \textbf{\bibinfo{volume}{58}},
  \bibinfo{pages}{1593} (\bibinfo{year}{1987}).

\bibitem[{\citenamefont{Anandan}(1992)}]{anandan92:_geomet_phase}
\bibinfo{author}{\bibfnamefont{J.}~\bibnamefont{Anandan}},
  \bibinfo{journal}{Nature} \textbf{\bibinfo{volume}{350}},
  \bibinfo{pages}{307} (\bibinfo{year}{1992}).

\bibitem[{\citenamefont{Samuel and
  Bhandari}(1988)}]{samuel88:_gener_settin_for_berry_phase}
\bibinfo{author}{\bibfnamefont{J.}~\bibnamefont{Samuel}} \bibnamefont{and}
  \bibinfo{author}{\bibfnamefont{R.}~\bibnamefont{Bhandari}},
  \bibinfo{journal}{Phys. Rev. Lett.} \textbf{\bibinfo{volume}{60}},
  \bibinfo{pages}{2339} (\bibinfo{year}{1988}).

\bibitem[{\citenamefont{Morinaga et~al.}(2007)\citenamefont{Morinaga, Monma,
  Honda, and Kitano}}]{morinaga07:_berry_phase_for_noncy_rotat}
\bibinfo{author}{\bibfnamefont{A.}~\bibnamefont{Morinaga}},
  \bibinfo{author}{\bibfnamefont{A.}~\bibnamefont{Monma}},
  \bibinfo{author}{\bibfnamefont{K.}~\bibnamefont{Honda}}, \bibnamefont{and}
  \bibinfo{author}{\bibfnamefont{M.}~\bibnamefont{Kitano}},
  \bibinfo{journal}{Phys. Rev. A} \textbf{\bibinfo{volume}{76}},
  \bibinfo{pages}{052109} (\bibinfo{year}{2007}).

\bibitem[{\citenamefont{Uhlmann}(1986)}]{uhlmann86:_paral_trans_and_quant_holo%
n}
\bibinfo{author}{\bibfnamefont{A.}~\bibnamefont{Uhlmann}},
  \bibinfo{journal}{Rep. Math. Phys.} \textbf{\bibinfo{volume}{24}},
  \bibinfo{pages}{229} (\bibinfo{year}{1986}).

\bibitem[{\citenamefont{Sj\"{o}qvist et~al.}(2000)\citenamefont{Sj\"{o}qvist,
  Pati, Ekert, Anandan, Ericsson, Oi, and
  Vedral}}]{sjoqvist00:_geomet_phases_for_mixed_states_in_intery}
\bibinfo{author}{\bibfnamefont{E.}~\bibnamefont{Sj\"{o}qvist}},
  \bibinfo{author}{\bibfnamefont{A.~K.} \bibnamefont{Pati}},
  \bibinfo{author}{\bibfnamefont{A.}~\bibnamefont{Ekert}},
  \bibinfo{author}{\bibfnamefont{J.~S.} \bibnamefont{Anandan}},
  \bibinfo{author}{\bibfnamefont{M.}~\bibnamefont{Ericsson}},
  \bibinfo{author}{\bibfnamefont{D.~K.~L.} \bibnamefont{Oi}}, \bibnamefont{and}
  \bibinfo{author}{\bibfnamefont{V.}~\bibnamefont{Vedral}},
  \bibinfo{journal}{Phys. Rev. Lett.} \textbf{\bibinfo{volume}{85}},
  \bibinfo{pages}{2845} (\bibinfo{year}{2000}).

\bibitem[{\citenamefont{Pancharatnam}(1956)}]{pancharatnam56:_proc}
\bibinfo{author}{\bibfnamefont{S.}~\bibnamefont{Pancharatnam}},
  \bibinfo{journal}{Proc. Ind. Acad. Sci. A} \textbf{\bibinfo{volume}{44}},
  \bibinfo{pages}{247} (\bibinfo{year}{1956}).

\bibitem[{\citenamefont{Berry}(1987)}]{berry87:_adiab_phase_and_panch_phase}
\bibinfo{author}{\bibfnamefont{M.~V.} \bibnamefont{Berry}},
  \bibinfo{journal}{J. Mod. Opt.} \textbf{\bibinfo{volume}{34}},
  \bibinfo{pages}{1401} (\bibinfo{year}{1987}).

\bibitem[{\citenamefont{Mukunda}(1993)}]{mukunda93:_quant_kinem_approac_to_geo%
met_phase}
\bibinfo{author}{\bibfnamefont{N.}~\bibnamefont{Mukunda}},
  \bibinfo{journal}{Ann. Phys.} \textbf{\bibinfo{volume}{228}},
  \bibinfo{pages}{205} (\bibinfo{year}{1993}).

\bibitem[{\citenamefont{Tomita and
  Chiao}(1986)}]{tomita86:_obser_of_berry_topol_phase}
\bibinfo{author}{\bibfnamefont{A.}~\bibnamefont{Tomita}} \bibnamefont{and}
  \bibinfo{author}{\bibfnamefont{R.~Y.} \bibnamefont{Chiao}},
  \bibinfo{journal}{Phys. Rev. Lett.} \textbf{\bibinfo{volume}{57}},
  \bibinfo{pages}{937} (\bibinfo{year}{1986}).

\bibitem[{\citenamefont{Simon et~al.}(1988)\citenamefont{Simon, Kimble, and
  Sudarshan}}]{simon88:_evolv_geomet_phase_and_its}
\bibinfo{author}{\bibfnamefont{R.}~\bibnamefont{Simon}},
  \bibinfo{author}{\bibfnamefont{H.~J.} \bibnamefont{Kimble}},
  \bibnamefont{and} \bibinfo{author}{\bibfnamefont{E.~C.~G.}
  \bibnamefont{Sudarshan}}, \bibinfo{journal}{Phys. Rev. Lett.}
  \textbf{\bibinfo{volume}{61}}, \bibinfo{pages}{19} (\bibinfo{year}{1988}).

\bibitem[{\citenamefont{Chiao et~al.}(1988)\citenamefont{Chiao, Antaramian,
  Ganga, Jiao, Wilkinson, and Nathel}}]{chiao88:_obser_of_topol_phase_by}
\bibinfo{author}{\bibfnamefont{R.~Y.} \bibnamefont{Chiao}},
  \bibinfo{author}{\bibfnamefont{A.}~\bibnamefont{Antaramian}},
  \bibinfo{author}{\bibfnamefont{K.~M.} \bibnamefont{Ganga}},
  \bibinfo{author}{\bibfnamefont{H.}~\bibnamefont{Jiao}},
  \bibinfo{author}{\bibfnamefont{S.~R.} \bibnamefont{Wilkinson}},
  \bibnamefont{and} \bibinfo{author}{\bibfnamefont{H.}~\bibnamefont{Nathel}},
  \bibinfo{journal}{Phys. Rev. Lett.} \textbf{\bibinfo{volume}{60}},
  \bibinfo{pages}{1214} (\bibinfo{year}{1988}).

\bibitem[{\citenamefont{Kwiat and
  Chiao}(1991)}]{kwiat91:_obser_of_noncl_berry_phase_for_photon}
\bibinfo{author}{\bibfnamefont{P.~G.} \bibnamefont{Kwiat}} \bibnamefont{and}
  \bibinfo{author}{\bibfnamefont{R.~Y.} \bibnamefont{Chiao}},
  \bibinfo{journal}{Phys. Rev. Lett.} \textbf{\bibinfo{volume}{66}},
  \bibinfo{pages}{588} (\bibinfo{year}{1991}).

\bibitem[{\citenamefont{Wagh and
  Rakhecha}(1995{\natexlab{a}})}]{wagh95:_measur_panch_phase}
\bibinfo{author}{\bibfnamefont{A.~G.} \bibnamefont{Wagh}} \bibnamefont{and}
  \bibinfo{author}{\bibfnamefont{V.~C.} \bibnamefont{Rakhecha}},
  \bibinfo{journal}{Phys. Lett. A} \textbf{\bibinfo{volume}{197}},
  \bibinfo{pages}{107} (\bibinfo{year}{1995}{\natexlab{a}}).

\bibitem[{\citenamefont{Wagh and
  Rakhecha}(1995{\natexlab{b}})}]{wagh95:_measur_panch_phase2}
\bibinfo{author}{\bibfnamefont{A.~G.} \bibnamefont{Wagh}} \bibnamefont{and}
  \bibinfo{author}{\bibfnamefont{V.~C.} \bibnamefont{Rakhecha}},
  \bibinfo{journal}{Phys. Lett. A} \textbf{\bibinfo{volume}{197}},
  \bibinfo{pages}{112} (\bibinfo{year}{1995}{\natexlab{b}}).

\bibitem[{\citenamefont{Loredo et~al.}(2009)\citenamefont{Loredo, Ort\'{i}z,
  Weing\"{a}rtner, and DeZela}}]{loredo09:_measur_of_panch_phase_by}
\bibinfo{author}{\bibfnamefont{J.~C.} \bibnamefont{Loredo}},
  \bibinfo{author}{\bibfnamefont{O.}~\bibnamefont{Ort\'{i}z}},
  \bibinfo{author}{\bibfnamefont{R.}~\bibnamefont{Weing\"{a}rtner}},
  \bibnamefont{and} \bibinfo{author}{\bibfnamefont{F.}~\bibnamefont{DeZela}},
  \bibinfo{journal}{Phys. Rev. A} \textbf{\bibinfo{volume}{80}},
  \bibinfo{pages}{012113} (\bibinfo{year}{2009}).

\bibitem[{\citenamefont{Kobayashi et~al.}(2011)\citenamefont{Kobayashi, Tamate,
  Nakanishi, Sugiyama, and
  Kitano}}]{kobayashi11:_obser_of_geomet_phases_in_quant_eraser}
\bibinfo{author}{\bibfnamefont{H.}~\bibnamefont{Kobayashi}},
  \bibinfo{author}{\bibfnamefont{S.}~\bibnamefont{Tamate}},
  \bibinfo{author}{\bibfnamefont{T.}~\bibnamefont{Nakanishi}},
  \bibinfo{author}{\bibfnamefont{K.}~\bibnamefont{Sugiyama}}, \bibnamefont{and}
  \bibinfo{author}{\bibfnamefont{M.}~\bibnamefont{Kitano}},
  \bibinfo{journal}{J. Phys. Soc. Jpn.} \textbf{\bibinfo{volume}{80}},
  \bibinfo{pages}{034401} (\bibinfo{year}{2011}).

\bibitem[{\citenamefont{Schmitzer et~al.}(1993)\citenamefont{Schmitzer, Klein,
  and Dultz}}]{schmitzer93:_nonlin_of_panch_topol_phase}
\bibinfo{author}{\bibfnamefont{H.}~\bibnamefont{Schmitzer}},
  \bibinfo{author}{\bibfnamefont{S.}~\bibnamefont{Klein}}, \bibnamefont{and}
  \bibinfo{author}{\bibfnamefont{W.}~\bibnamefont{Dultz}},
  \bibinfo{journal}{Phys. Rev. Lett.} \textbf{\bibinfo{volume}{71}},
  \bibinfo{pages}{1530} (\bibinfo{year}{1993}).

\bibitem[{\citenamefont{Tewari et~al.}(1995)\citenamefont{Tewari, Ashoka, and
  Ramana}}]{tewari95:_four_arm_sagnac_inter_switc}
\bibinfo{author}{\bibfnamefont{S.~P.} \bibnamefont{Tewari}},
  \bibinfo{author}{\bibfnamefont{V.~S.} \bibnamefont{Ashoka}},
  \bibnamefont{and} \bibinfo{author}{\bibfnamefont{M.~S.}
  \bibnamefont{Ramana}}, \bibinfo{journal}{Opt. Commun.}
  \textbf{\bibinfo{volume}{120}}, \bibinfo{pages}{235} (\bibinfo{year}{1995}).

\bibitem[{\citenamefont{Bhandari}(1997)}]{bhandari97:_polar_of_light_and_topol%
_phases}
\bibinfo{author}{\bibfnamefont{R.}~\bibnamefont{Bhandari}},
  \bibinfo{journal}{Phys. Rep.} \textbf{\bibinfo{volume}{281}},
  \bibinfo{pages}{1} (\bibinfo{year}{1997}).

\bibitem[{\citenamefont{Hils et~al.}(1999)\citenamefont{Hils, Dultz, and
  Martienssen}}]{hils99:_nonlin_of_pnach_geomet_phase}
\bibinfo{author}{\bibfnamefont{B.}~\bibnamefont{Hils}},
  \bibinfo{author}{\bibfnamefont{W.}~\bibnamefont{Dultz}}, \bibnamefont{and}
  \bibinfo{author}{\bibfnamefont{W.}~\bibnamefont{Martienssen}},
  \bibinfo{journal}{Phys. Rev. E} \textbf{\bibinfo{volume}{60}},
  \bibinfo{pages}{2322} (\bibinfo{year}{1999}).

\bibitem[{\citenamefont{Li et~al.}(1999)\citenamefont{Li, Gong, Gao, and
  Chen}}]{li99:_exper_obser_of_nonlin_of}
\bibinfo{author}{\bibfnamefont{Q.}~\bibnamefont{Li}},
  \bibinfo{author}{\bibfnamefont{L.}~\bibnamefont{Gong}},
  \bibinfo{author}{\bibfnamefont{Y.}~\bibnamefont{Gao}}, \bibnamefont{and}
  \bibinfo{author}{\bibfnamefont{Y.}~\bibnamefont{Chen}},
  \bibinfo{journal}{Opt. Commun.} \textbf{\bibinfo{volume}{169}},
  \bibinfo{pages}{17} (\bibinfo{year}{1999}).

\bibitem[{\citenamefont{Tamate et~al.}(2009)\citenamefont{Tamate, Kobayashi,
  Nakanishi, Sugiyama, and Kitano}}]{tamate09:_geomet_aspec_of_weak_measur}
\bibinfo{author}{\bibfnamefont{S.}~\bibnamefont{Tamate}},
  \bibinfo{author}{\bibfnamefont{H.}~\bibnamefont{Kobayashi}},
  \bibinfo{author}{\bibfnamefont{T.}~\bibnamefont{Nakanishi}},
  \bibinfo{author}{\bibfnamefont{K.}~\bibnamefont{Sugiyama}}, \bibnamefont{and}
  \bibinfo{author}{\bibfnamefont{M.}~\bibnamefont{Kitano}},
  \bibinfo{journal}{New J. Phys.} \textbf{\bibinfo{volume}{11}},
  \bibinfo{pages}{093025} (\bibinfo{year}{2009}).

\bibitem[{\citenamefont{Schmitzer et~al.}(1991)\citenamefont{Schmitzer, Klein,
  and Dultz}}]{schmitzer91:_optic_switc_based_panch_topol_phase}
\bibinfo{author}{\bibfnamefont{H.}~\bibnamefont{Schmitzer}},
  \bibinfo{author}{\bibfnamefont{S.}~\bibnamefont{Klein}}, \bibnamefont{and}
  \bibinfo{author}{\bibfnamefont{W.}~\bibnamefont{Dultz}},
  \bibinfo{journal}{Physica B} \textbf{\bibinfo{volume}{175}},
  \bibinfo{pages}{148} (\bibinfo{year}{1991}).

\bibitem[{\citenamefont{Klyshko}(1989)}]{klyshko89:_berry_phase_in_multip_expe%
r}
\bibinfo{author}{\bibfnamefont{D.~N.} \bibnamefont{Klyshko}},
  \bibinfo{journal}{Phys. Lett. A} \textbf{\bibinfo{volume}{140}},
  \bibinfo{pages}{19} (\bibinfo{year}{1989}).

\bibitem[{\citenamefont{Brendel et~al.}(1995)\citenamefont{Brendel, Dultz, and
  Martienssen}}]{brendel95:_geomet_phases_in_two_photon_inter_exper}
\bibinfo{author}{\bibfnamefont{J.}~\bibnamefont{Brendel}},
  \bibinfo{author}{\bibfnamefont{W.}~\bibnamefont{Dultz}}, \bibnamefont{and}
  \bibinfo{author}{\bibfnamefont{W.}~\bibnamefont{Martienssen}},
  \bibinfo{journal}{Phys. Rev. A} \textbf{\bibinfo{volume}{52}},
  \bibinfo{pages}{2551} (\bibinfo{year}{1995}).

\bibitem[{\citenamefont{Sj\"oqvist}(2000)}]{sjoeqvist00:_geomet_phase_for_enta%
n_spin_pairs}
\bibinfo{author}{\bibfnamefont{E.}~\bibnamefont{Sj\"oqvist}},
  \bibinfo{journal}{Phys. Rev. A} \textbf{\bibinfo{volume}{62}},
  \bibinfo{pages}{022109} (\bibinfo{year}{2000}).

\bibitem[{\citenamefont{Hessmo and
  Sj\"oqvist}(2000)}]{hessmo00:_quant_phase_for_nonmax_entan_photon}
\bibinfo{author}{\bibfnamefont{B.}~\bibnamefont{Hessmo}} \bibnamefont{and}
  \bibinfo{author}{\bibfnamefont{E.}~\bibnamefont{Sj\"oqvist}},
  \bibinfo{journal}{Phys. Rev. A} \textbf{\bibinfo{volume}{62}},
  \bibinfo{pages}{062301} (\bibinfo{year}{2000}).

\bibitem[{\citenamefont{Ge and
  Wadati}(2005)}]{ge05:_geomet_phase_of_entan_spin}
\bibinfo{author}{\bibfnamefont{X.-Y.} \bibnamefont{Ge}} \bibnamefont{and}
  \bibinfo{author}{\bibfnamefont{M.}~\bibnamefont{Wadati}},
  \bibinfo{journal}{Phys. Rev. A} \textbf{\bibinfo{volume}{72}},
  \bibinfo{pages}{052101} (\bibinfo{year}{2005}).

\bibitem[{\citenamefont{Williamson and
  Vedral}(2007)}]{williamson07:_compos_geomet_phase_for_multip_entan_states}
\bibinfo{author}{\bibfnamefont{M.~S.} \bibnamefont{Williamson}}
  \bibnamefont{and} \bibinfo{author}{\bibfnamefont{V.}~\bibnamefont{Vedral}},
  \bibinfo{journal}{Phys. Rev. A} \textbf{\bibinfo{volume}{76}},
  \bibinfo{pages}{032115} (\bibinfo{year}{2007}).

\bibitem[{\citenamefont{Aharonov et~al.}(1988)\citenamefont{Aharonov, Albert,
  and Vaidman}}]{aharonov88:_how_resul_of_measur_of}
\bibinfo{author}{\bibfnamefont{Y.}~\bibnamefont{Aharonov}},
  \bibinfo{author}{\bibfnamefont{D.~Z.} \bibnamefont{Albert}},
  \bibnamefont{and} \bibinfo{author}{\bibfnamefont{L.}~\bibnamefont{Vaidman}},
  \bibinfo{journal}{Phys. Rev. Lett.} \textbf{\bibinfo{volume}{60}},
  \bibinfo{pages}{1351} (\bibinfo{year}{1988}).

\bibitem[{\citenamefont{Hosten and
  Kwiat}(2008)}]{hosten08:_obser_of_spin_hall_effec}
\bibinfo{author}{\bibfnamefont{O.}~\bibnamefont{Hosten}} \bibnamefont{and}
  \bibinfo{author}{\bibfnamefont{P.}~\bibnamefont{Kwiat}},
  \bibinfo{journal}{Science} \textbf{\bibinfo{volume}{319}},
  \bibinfo{pages}{787} (\bibinfo{year}{2008}).

\bibitem[{\citenamefont{Dixon et~al.}(2009)\citenamefont{Dixon, Starling,
  Jordan, and Howell}}]{dixon09:_ultras_beam_deflec_measur_via}
\bibinfo{author}{\bibfnamefont{P.~B.} \bibnamefont{Dixon}},
  \bibinfo{author}{\bibfnamefont{D.~J.} \bibnamefont{Starling}},
  \bibinfo{author}{\bibfnamefont{A.~N.} \bibnamefont{Jordan}},
  \bibnamefont{and} \bibinfo{author}{\bibfnamefont{J.~C.}
  \bibnamefont{Howell}}, \bibinfo{journal}{Phys. Rev. Lett.}
  \textbf{\bibinfo{volume}{102}}, \bibinfo{pages}{173601}
  (\bibinfo{year}{2009}).

\bibitem[{\citenamefont{Starling et~al.}(2009)\citenamefont{Starling, Dixon,
  Jordan, and Howell}}]{starling09:_optim_signal_to_noise_ratio}
\bibinfo{author}{\bibfnamefont{D.~J.} \bibnamefont{Starling}},
  \bibinfo{author}{\bibfnamefont{P.~B.} \bibnamefont{Dixon}},
  \bibinfo{author}{\bibfnamefont{A.~N.} \bibnamefont{Jordan}},
  \bibnamefont{and} \bibinfo{author}{\bibfnamefont{J.~C.}
  \bibnamefont{Howell}}, \bibinfo{journal}{Phys. Rev. A}
  \textbf{\bibinfo{volume}{80}}, \bibinfo{pages}{041803}
  (\bibinfo{year}{2009}).

\bibitem[{\citenamefont{Feizpour et~al.}(2010)\citenamefont{Feizpour, Xing, and
  Steinberg}}]{feizpour10:_weak_measur_amplif_of_singl_photon_nonlin}
\bibinfo{author}{\bibfnamefont{A.}~\bibnamefont{Feizpour}},
  \bibinfo{author}{\bibfnamefont{X.}~\bibnamefont{Xing}}, \bibnamefont{and}
  \bibinfo{author}{\bibfnamefont{A.~M.} \bibnamefont{Steinberg}},
  \bibinfo{journal}{arXiv:1101.0199}  (\bibinfo{year}{2010}).

\bibitem[{\citenamefont{Aravind}(1992)}]{aravind92:_simpl_proof_of_panch_theor}
\bibinfo{author}{\bibfnamefont{P.~K.} \bibnamefont{Aravind}},
  \bibinfo{journal}{Opt. Commun.} \textbf{\bibinfo{volume}{94}},
  \bibinfo{pages}{191} (\bibinfo{year}{1992}).

\bibitem[{\citenamefont{Sanders}(1989)}]{sanders89:_quant_dynam_of_nonlin_rota%
t}
\bibinfo{author}{\bibfnamefont{B.~C.} \bibnamefont{Sanders}},
  \bibinfo{journal}{Phys. Rev. A} \textbf{\bibinfo{volume}{40}},
  \bibinfo{pages}{2417} (\bibinfo{year}{1989}).

\bibitem[{\citenamefont{Bollinger et~al.}(1996)\citenamefont{Bollinger, Itano,
  Wineland, and
  Heinzen}}]{bollinger96:_optim_frequen_measur_with_maxim_correl_states}
\bibinfo{author}{\bibfnamefont{J.~J.~.} \bibnamefont{Bollinger}},
  \bibinfo{author}{\bibfnamefont{W.~M.} \bibnamefont{Itano}},
  \bibinfo{author}{\bibfnamefont{D.~J.} \bibnamefont{Wineland}},
  \bibnamefont{and} \bibinfo{author}{\bibfnamefont{D.~J.}
  \bibnamefont{Heinzen}}, \bibinfo{journal}{Phys. Rev. A}
  \textbf{\bibinfo{volume}{54}}, \bibinfo{pages}{R4649} (\bibinfo{year}{1996}).

\bibitem[{\citenamefont{Boto et~al.}(2000)\citenamefont{Boto, Kok, Abrams,
  Braunstein, Williams, and Dowling}}]{boto00:_quant_inter_optic_lithog}
\bibinfo{author}{\bibfnamefont{A.~N.} \bibnamefont{Boto}},
  \bibinfo{author}{\bibfnamefont{P.}~\bibnamefont{Kok}},
  \bibinfo{author}{\bibfnamefont{D.~S.} \bibnamefont{Abrams}},
  \bibinfo{author}{\bibfnamefont{S.~L.} \bibnamefont{Braunstein}},
  \bibinfo{author}{\bibfnamefont{C.~P.} \bibnamefont{Williams}},
  \bibnamefont{and} \bibinfo{author}{\bibfnamefont{J.~P.}
  \bibnamefont{Dowling}}, \bibinfo{journal}{Phys. Rev. Lett.}
  \textbf{\bibinfo{volume}{85}}, \bibinfo{pages}{2733} (\bibinfo{year}{2000}).

\bibitem[{\citenamefont{Edamatsu et~al.}(2002)\citenamefont{Edamatsu, Shimizu,
  and Itoh}}]{edamatsu02:_measur_of_photon_de_brogl}
\bibinfo{author}{\bibfnamefont{K.}~\bibnamefont{Edamatsu}},
  \bibinfo{author}{\bibfnamefont{R.}~\bibnamefont{Shimizu}}, \bibnamefont{and}
  \bibinfo{author}{\bibfnamefont{T.}~\bibnamefont{Itoh}},
  \bibinfo{journal}{Phys. Rev. Lett.} \textbf{\bibinfo{volume}{89}},
  \bibinfo{pages}{213601} (\bibinfo{year}{2002}).

\bibitem[{\citenamefont{Walther et~al.}(2004)\citenamefont{Walther, Pan,
  Aspelmeyer, Ursin, Gasparoni, and
  Zeilinger}}]{philip04:_de_brogl_wavel_of_non}
\bibinfo{author}{\bibfnamefont{P.}~\bibnamefont{Walther}},
  \bibinfo{author}{\bibfnamefont{J.-W.} \bibnamefont{Pan}},
  \bibinfo{author}{\bibfnamefont{M.}~\bibnamefont{Aspelmeyer}},
  \bibinfo{author}{\bibfnamefont{R.}~\bibnamefont{Ursin}},
  \bibinfo{author}{\bibfnamefont{S.}~\bibnamefont{Gasparoni}},
  \bibnamefont{and}
  \bibinfo{author}{\bibfnamefont{A.}~\bibnamefont{Zeilinger}},
  \bibinfo{journal}{Nature} \textbf{\bibinfo{volume}{429}},
  \bibinfo{pages}{158} (\bibinfo{year}{2004}).

\bibitem[{\citenamefont{Nagata et~al.}(2007)\citenamefont{Nagata, Okamoto,
  O'Brien, Sasaki, and Takeuchi}}]{nagata07:_beatin_stand_quant_limit_with}
\bibinfo{author}{\bibfnamefont{T.}~\bibnamefont{Nagata}},
  \bibinfo{author}{\bibfnamefont{R.}~\bibnamefont{Okamoto}},
  \bibinfo{author}{\bibfnamefont{J.~L.} \bibnamefont{O'Brien}},
  \bibinfo{author}{\bibfnamefont{K.}~\bibnamefont{Sasaki}}, \bibnamefont{and}
  \bibinfo{author}{\bibfnamefont{S.}~\bibnamefont{Takeuchi}},
  \bibinfo{journal}{Science} \textbf{\bibinfo{volume}{316}},
  \bibinfo{pages}{726} (\bibinfo{year}{2007}).

\bibitem[{\citenamefont{Dowling}(2008)}]{dowling08:_quant_optic_metrol}
\bibinfo{author}{\bibfnamefont{J.~P.} \bibnamefont{Dowling}},
  \bibinfo{journal}{Contemp. Phys.} \textbf{\bibinfo{volume}{49}},
  \bibinfo{pages}{125} (\bibinfo{year}{2008}).

\bibitem[{\citenamefont{Hong et~al.}(1987)\citenamefont{Hong, Ou, and
  Mandel}}]{hong87:_measur_of_subpic_time_inter}
\bibinfo{author}{\bibfnamefont{C.~K.} \bibnamefont{Hong}},
  \bibinfo{author}{\bibfnamefont{Z.~Y.} \bibnamefont{Ou}}, \bibnamefont{and}
  \bibinfo{author}{\bibfnamefont{L.}~\bibnamefont{Mandel}},
  \bibinfo{journal}{Phys. Rev. Lett.} \textbf{\bibinfo{volume}{59}},
  \bibinfo{pages}{2044} (\bibinfo{year}{1987}).

\bibitem[{\citenamefont{Branning et~al.}(2009)\citenamefont{Branning, Bhandari,
  and Beck}}]{branning09:_low_cost_coinc_count_elect}
\bibinfo{author}{\bibfnamefont{D.}~\bibnamefont{Branning}},
  \bibinfo{author}{\bibfnamefont{S.}~\bibnamefont{Bhandari}}, \bibnamefont{and}
  \bibinfo{author}{\bibfnamefont{M.}~\bibnamefont{Beck}}, \bibinfo{journal}{Am.
  J. Phys.} \textbf{\bibinfo{volume}{77}}, \bibinfo{pages}{667}
  (\bibinfo{year}{2009}).

\end{thebibliography}
\end{document}